\documentclass[preprint,prd,amsmath,amssymb,amsthm,nofootinbib]{revtex4}
\usepackage{amsmath}
\usepackage{ulem}
\usepackage{xcolor}
\usepackage{multirow}
\usepackage{array}
\usepackage{booktabs}
\usepackage{threeparttable}
\usepackage[mathscr]{euscript}
\makeatletter

\newcommand{\Rmnum}[1]{\expandafter\@slowromancap\romannumeral #1@}
\makeatother
\usepackage{graphicx}% Include figure files
\usepackage{subfigure}
\usepackage{tikz}
\usetikzlibrary{positioning}
\usepackage{bm}% bold math
\usepackage{amsmath,amssymb,amsthm}
\usepackage[colorlinks=true,linkcolor=red]{hyperref}
\newcommand{\bea}{\begin{eqnarray}}
\newcommand{\eea}{\end{eqnarray}}
\newcommand{\beq}{\begin{equation}}
\newcommand{\eeq}{\end{equation}}
\newcommand{\nn}{\nonumber}
\def\/{\over}

\begin{document}
\title{Entanglement dynamics for Unruh-DeWitt detectors interacting with massive scalar fields: The Unruh and anti-Unruh effects}

\author{Yuebing Zhou$^{1,2}$,
Jiawei Hu$^{1}$\footnote{Corresponding author: jwhu@hunnu.edu.cn} and
Hongwei Yu$^{1}$\footnote{Corresponding author: hwyu@hunnu.edu.cn }}

\affiliation{
$^{1}$Department of Physics and Synergetic Innovation Center for Quantum Effects and Applications, Hunan Normal University, 36 Lushan Rd., Changsha, Hunan 410081, China\\
$^2$Department of Physics, Huaihua University, 180 Huaidong Rd., Huaihua, Hunan 418008, China}

\begin{abstract}

We study, in the framework of open quantum systems, the entanglement dynamics for a quantum system composed of two uniformly accelerated Unruh-Dewitt detectors interacting with a bath of massive scalar fields in the Minkowski vacuum. We find that the entanglement evolution  for the quantum system coupled with massive fields is always slower compared with that of the one coupled with massless fields, and this time-delay effect brought about by the field being massive can however be counteracted by a large enough acceleration,
in contrast to the case of a static quantum system in a thermal bath, where this time delay is not  affected by the temperature.
Remarkably, the maximal concurrence of the quantum system generated during evolution may increase  with acceleration for any inter-detector separation while that for static ones in a thermal bath decreases monotonically with temperature, and this can be considered as an anti-Unruh effect in terms of the entanglement generated.

\end{abstract}

\maketitle

\section{Introduction}
Quantum field theory predicts that a uniformly accelerated observer perceives the vacuum of an inertial observer as a thermal bath at a temperature proportional to its proper acceleration, which is known as the Unruh effect~\cite{Fulling1973,W. G. Unruh,Davies1975,Crispino2008}.
A widely used model of uniformly accelerated observers is the Unruh-DeWitt detector,  which is typically modeled as a point-like two-level quantum system coupled with  fluctuating vacuum quantum fields \cite{W. G. Unruh,dewitt}.
In fact, the thermal bath perceived by the uniformly accelerated observer  can be considered as the Rindler thermal bath, which may not necessarily be same as the Minkowski thermal bath seen by an inertial observer.
Many studies have been committed to comparing the different behaviors between quantum systems immersed in these two kinds of thermal baths from different physical aspects, e.g. the  spontaneous emission rates \cite{Takagi1986,J. Audretsch1994,Lu2005,Zhu2006}, the Lamb shifts \cite{J. Audretsch1995,Passante1998}, the resonance interactions \cite{Rizzuto2016, Wentingzhou2016} and the Casimir-Polder interactions \cite{Rizzuto2007,Zhu2010,Passante2014}. It has been shown that, only in some  special situations, the behaviors  of a single Unruh-DeWitt detector in the two kinds of thermal baths are equivalent, such as in the case of massless scalar fields in a free spacetime \cite{J. Audretsch1994,J. Audretsch1995}.

Recently, the influences of  environment on the entanglement dynamics of an open quantum system have been extensively studied, such as those leading to the environment-induced entanglement sudden death \cite{T. Yu3,  T. Yu}, entanglement revival \cite{Ficek2} and entanglement creation \cite{Braun,Kim,Schneider,Basharov,Jakobczyk,Reznik,Piani,Z. Ficek, R. Tanas,esb-1,esb-2,esb-3,esb-4}.
Therefore, a question naturally arises as to how the entanglement dynamics of a quantum system composed of two Unruh-DeWitt detectors will be affected by acceleration, and how will it be different from that of a static one in a thermal bath in the Minkowski spacetime at the Unruh temperature related to acceleration.
In Ref.~\cite{Benatti}, the entanglement generation of two uniformly accelerated Unruh-DeWitt detectors coupled with fluctuating massless scalar fields in the Minkowski vacuum with a vanishing separation has been studied, and it has been shown that the asymptotic entanglement is exactly the same as that immersed in a thermal bath at the Unruh temperature.
However, in more general cases, e.g. when the separation between the detectors is nonzero \cite{Hu,S. Cheng1,Yang}, in the presence of a boundary \cite{yu-prd-07,S. Cheng1}, and for the quantum systems coupled with  different kinds of quantum fields, such as electromagnetic fields \cite{Yang,S. Cheng1},
the differences between the entanglement dynamics of an accelerated quantum system and that of a static one in a thermal bath show up.

When an Unruh-DeWitt detector is coupled with massive scalar fields, transitions among different eigenstates for an accelerated detector can still occur even when the mass of the field is greater than the energy level spacing of the detector \cite{Takagi1986,Lu2005,Crispino2008,Y. B. Zhou}, which is impossible for a static one in a thermal bath. Furthermore, it has been shown in  Ref. \cite{Y. B. Zhou} that the mass of the field will bring a gray factor related to acceleration to the transition rate of an Unruh-DeWitt detector, which results in the fact that the transition rate may decrease with acceleration. Later, this  phenomenon is named as the anti-Unruh effect \cite{W. Brenna,Anti-Unruh2016}.
Recently, the entanglement dynamics of two static detectors coupled with massive scalar fields has been studied in Ref. \cite{Zhou2020}. It has been found that, compared with the massless field case, the evolution of entanglement is slower and the region of spatial separation between the detectors within which entanglement can be generated is significantly enlarged.
This means that it is possible to achieve long-distance entanglement generation and long-lived entanglement.
Now, a natural question is, will there be essential differences between the behaviors of two uniformly accelerated Unruh-DeWitt detectors coupled with massive fields in the Minkowski vacuum and that of a static one in a thermal bath in terms of entanglement dynamics?
In particular, will there be anti-Unruh phenomena, e.g. the entanglement generated for accelerated detectors increases with acceleration, while that for static ones in a thermal bath decreases with temperature?
In the present paper, we study, in the framework of open quantum systems, the entanglement dynamics for a uniformly accelerated quantum system composed of two Unruh-DeWitt detectors interacting with a bath of fluctutating massive scalar fields in the Minkowski vacuum. Hereafter natural units with $\hbar=c=k_B=1$ are used  unless specified, where $c$ is the speed of light, $\hbar$  the reduced Planck constant, and $k_B$  the  Boltzmann constant.

\section{The basic formalism}
We consider a quantum system composed of a pair of Unruh-DeWitt detectors in interaction with a bath of fluctuating massive scalar fields in the Minkowski vacuum.
The Hamiltonian of the total system takes the form
\begin{equation}
H=H_{S}+H_{F}+H_{I}.
\end{equation}
Here  $H_{S}$ denotes the Hamiltonian of the  quantum system, which can be generically written as
\begin{equation}
H_{S}=\frac{\omega}{2}R_{3}^{(1)}+\frac{\omega}{2}R_{3}^{(2)},
\end{equation}
where $\omega$ is the energy level spacing between the excited state $|1\rangle$ and the ground state $|0\rangle$ of the Unruh-Dewitt detector,
and $R_{\mu}^{(1)}=R_{\mu}\otimes R_{0}$, $R_{\mu}^{(2)}=R_{0}\otimes R_{\mu}$, with $R_{0}=|1\rangle\langle1|+|0\rangle\langle0|$,
$R_{1}=|0\rangle\langle1|+|1\rangle\langle0|$, $R_{2}=i(|0\rangle\langle1|-|1\rangle\langle0|)$, and $R_{3}=|1\rangle\langle1|-|0\rangle\langle0|$.
$H_{F}$ is the Hamiltonian of the massive scalar fields, the detail of which is not relevant here. The interaction Hamiltonian $H_{I}$ can be  written in the following general form as \cite {Benatti}
\begin{equation}
H_{I}=\varepsilon\sum_{\alpha=1}^{2}\sum_{\mu=0}^{3}R_{\mu}^{(\alpha)}\Phi_{\mu}(t_{\alpha},\mathbf{x_{\alpha}}),
\end{equation}
where $\Phi_{\mu}(t,\mathbf{x})$ is the field operator, and $\varepsilon$ is the coupling constant which is assumed to be small. Now we assume that the scalar fields can be expanded as
\begin{equation}\label{Phi}
\Phi_{\mu}(x)=\sum^N_{k=1}\,[\chi_\mu^k\phi^{(-)}_{k}(x) +
(\chi_\mu^k)^*\phi^{(+)}_{k}(x)]\;,
\end{equation}
where $\phi^{(\pm)}_{k}(x)$ are positive and negative energy field
operators relative to a set of $N$ independent massive scalar fields, and $\chi_\mu^k$ are the
complex coefficients~\cite{Benatti}.

We assume that initially the quantum system is  uncorrelated with the environment, i.e., the initial state of the total system can be written as $\rho_{\rm tot}(0)=\rho(0)\otimes|0\rangle\langle0|$,
where $|0\rangle$ is the vacuum state of the massive scalar fields, and $\rho(0)$ denotes the initial state of the quantum system. The density matrix of the total system satisfies the Liouville equation
\begin{equation}
\frac{\partial\rho_{\rm tot}(\tau)}{\partial \tau} = -i[H,\rho_{\rm tot}(\tau)].
\end{equation}
Under the Born-Markov approximation, the reduced density matrix of the quantum system $\rho(\tau)=\mathrm{Tr}_{F}[\rho_{\rm tot}(\tau)]$ can be described by the  Gorini-Kossakowski-Lindblad-Sudarshan (GKLS) master equation \cite{Kossakowski, Lindblad},
\begin{equation}\label{m}
\frac{\partial\rho(\tau)}{\partial \tau} = -i[H_{\rm eff},\rho(\tau)] +\mathcal{D}[\rho(\tau)].
\end{equation}
$H_{\rm eff}$ and $\mathcal{D}[\rho(\tau)]$ in the  equation above  are the effective Hamiltonian and the dissipative term respectively, whose explicit form can be written as
\bea
&&H_{\rm eff}=H_{S}-\frac{i}{2}\sum\limits_{\alpha,\varrho=1}^{2} \Big[H_{+}^{(\alpha\varrho)}R_{+}^{(\alpha)}R_{-}^{(\varrho)}+H_{-}^{(\alpha\varrho)}R_{-}^{(\alpha)}R_{+}^{(\varrho)}+H_{0}^{(\alpha\varrho)}R_{3}^{(\alpha)}R_{3}^{(\varrho)}\Big],\nonumber
\eea
and
\bea
\mathcal{D}[\rho(\tau)]&=&\frac{1}{2}\sum\limits_{\alpha,\varrho=1}^{2} \Big[D_{+}^{(\alpha\varrho)}\left(2R_{-}^{(\varrho)}\rho R_{+}^{(\alpha)}-\big\{R_{+}^{(\alpha)}
R_{-}^{(\varrho)},\rho\big\}\right)\nonumber\\
&&\;\;\;\;\;\;\;\;\;\;+D_{-}^{(\alpha\varrho)}\left(2R_{+}^{(\varrho)}\rho R_{-}^{(\alpha)}-\big\{R_{-}^{(\alpha)}
R_{+}^{(\varrho)},\rho\big\}\right)\nonumber\\
&&\;\;\;\;\;\;\;\;\;\;+D_{0}^{(\alpha\varrho)}\left(2R_{3}^{(\varrho)}\rho R_{3}^{(\alpha)}-\big\{R_{3}^{(\alpha)}
R_{3}^{(\varrho)},\rho\big\}\right)\Big],
\eea
where $R_{\pm}^{(1)}=R_{\pm}\otimes R_{0}$ and $R_{\pm}^{(2)}=R_{0}\otimes R_{\pm}$ with $R_{-}=|0\rangle\langle 1|$ and $R_{+}=|1\rangle\langle 0|$.
Here, $D_{\kappa}^{(\alpha\varrho)}$ and $H_{\kappa}^{(\alpha\varrho)}$  ($\kappa=+,-,0$) are determined by the Fourier and Hilbert transforms of the
field correlation functions
\begin{eqnarray}\label{C+chi}
&&G^{(\alpha\varrho)}_{ij}(\Delta\tau)=\big\langle\Phi_{i}\big(t_{\alpha}(\tau),\mathbf{x}_{\alpha}(\tau)\big)\Phi_{j}\big(t_{\varrho}(\tau'),\mathbf{x}_{\varrho}(\tau')\big)\big\rangle.
\end{eqnarray}
Note here that, $\langle\cdot\cdot\cdot\rangle$ denotes the expectation value with respect to a certain state of the quantum field.
Moreover, since  the environment  perceived by the quantum system is stationary,  the correlation functions of the fields Eq. \eqref{C+chi} are  functions of $\Delta\tau=\tau-\tau'$. Recalling that the field variables $\phi^{(\pm)}_{k}$
in  Eq. \eqref{Phi} are all independent, one finds
\bea
G^{(\alpha\varrho)}_{ij}(\Delta\tau)=\sum^N_{k=1}\,\chi^k_i(\chi^k_j)^*G^{(\alpha\varrho)}(\Delta\tau),
\eea
where $G^{(\alpha\varrho)} (\Delta\tau)$ is the standard Wightman function for a single scalar field. When the scalar field is in the Minkowski vacuum, the explicit form of  $G^{(\alpha\varrho)} (\Delta\tau)$ can be written as
\begin{eqnarray}\label{G}
&&G^{(\alpha\varrho)}(\Delta\tau)=-\frac{1}{4\pi^2}\frac{1}{[t_{\alpha}(\tau)-t_{\varrho}(\tau')-i \epsilon]^2-[\textbf{x}_{\alpha}(\tau)-\textbf{x}_{\varrho}(\tau')]^2}.
\end{eqnarray}
We define two functions ${\cal G}_{ij}^{(\alpha\varrho)}(x)$ and ${\cal K}_{ij}^{(\alpha\varrho)}(x)$ with the Fourier and Hilbert transformations of $G^{(\alpha\varrho)} (\Delta\tau)$ as
\begin{equation}
{\cal G}_{ij}^{(\alpha\varrho)}(x)=\varepsilon^2\int_{-\infty}^{\infty} dt \,
e^{i x t}\, G_{ij}^{(\alpha\varrho)}(t)\; , \label{fourierG}
\end{equation}
\begin{equation}
{\cal K}_{ij}^{(\alpha\varrho)}(x)=\varepsilon^2\int_{-\infty}^{\infty} dt \,
{\rm sign}(t)\, e^{i x t}\, G_{ij}^{(\alpha\varrho)}(t)=
\varepsilon^2\frac{P}{\pi i}\int_{-\infty}^{\infty} d\omega\ \frac{ {\cal
G}_{ij}^{(\alpha\varrho)}(\omega) }{\omega-x} \;, \label{kij}
\end{equation}
where $P$ denotes principal value. Then, the coefficient matrix $D_{\kappa}^{(\alpha\varrho)}$ can be written explicitly as
\begin{equation}
D_{\pm}^{(\alpha\varrho)}=\sum_{i,j=1}^{2}(\delta_{ij}\pm i \epsilon_{ij3}){\cal G}_{ij}^{(\alpha\varrho)}(\pm\omega),
\;\;\;\;D_{0}^{(\alpha\varrho)}={\cal G}_{33}^{(\alpha\varrho)}(0).\label{D}
\end{equation}
Similarly, the coefficients of $H_{\kappa}^{(\alpha\varrho)}$
can be obtained by replacing ${\cal
G}_{ij}^{(\alpha\varrho)}$ with ${\cal
K}_{ij}^{(\alpha\varrho)}$ in the above expressions.
For the sake of simplicity
, we now assume the coefficients $\chi^a_\mu$ satisfy
$\sum^N_{a=1}\,\chi^a_\mu(\chi^a_\nu)^*=\delta_{\mu\nu}$,
which, for example, can be fulfilled by assuming that the field components $\Phi_i (x)$ are
independent.
Then, $D_{\kappa}^{(\alpha\varrho)}$ can be written explicitly as
\bea\label{D2}
D_{\pm}^{(\alpha\varrho)}=2{\cal G}^{(\alpha\varrho)}(\pm\omega),
\;\;\;\;D_{0}^{(\alpha\varrho)}={\cal G}^{(\alpha\varrho)}(0),
\eea
where
\bea\label{mG}
{\cal G}^{(\alpha\varrho)}(x)=\varepsilon^2\int_{-\infty}^{\infty} dt \,
e^{i x t}\, G^{(\alpha\varrho)}(t)\; . \label{fourierG2}
\eea

In the following, we define a set of dimensionless physical parameters using $D_{\kappa}^{(\alpha\varrho)}$ to describe the  entanglement dynamics of the quantum system.

{\it 1. $\Omega$.}
The  factor $\Omega$ is defined as
\begin{eqnarray}\label{Omegad}
\Omega=\frac{\Gamma^{(\alpha\alpha)}}{\Gamma^{(\alpha\alpha)}\big|_{m=0}},
\end{eqnarray}
where $\Gamma^{(\alpha\varrho)}=D_{+}^{(\alpha\varrho)}-D_{-}^{(\alpha\varrho)}$. $\Omega$ represents the ratio of the  spontaneous emission rate of an Unruh-DeWitt detector coupled with massive scalar fields to that of the one coupled with massless scalar fields.
Here, the spontaneous emission rate of the Unruh-DeWitt detector coupled to massless scalar fields in the Minkowski vacuum is $\Gamma_0\equiv\Gamma^{(11)}\big|_{m=0}=\Gamma^{(22)}\big|_{m=0}=\varepsilon^2\omega/{\pi}$, with $m$ being the mass of the field.

In the acceleration case, we can obtain
\begin{eqnarray}\label{g}
\Omega_a=\frac{\sinh\left(\pi\omega/a\right)}{\pi\omega/a}
\frac{m^2}{a^2}\left[K_{1+i\omega/a}\left(m/a\right)K_{-1+i\omega/a}\left(m/a\right)
-K_{i\omega/a}^2\left(m/a\right)\right],
\end{eqnarray}
where $K_{\nu}(x)$ is the second type of modified Bessel function, and $a$ is the proper acceleration of the detectors. For the convenience of later discussions, $ \Omega_a $ is written as a function of dimensionless variables $m/\omega$ and $a/\omega$, i.e., $\Omega_a(m/\omega,a/\omega)$ (See Appendix \ref{the properties Omega} for the details of the properties of $\Omega_a$).
For comparison, for the thermal bath case,  $\Omega_{\beta}$ can be obtained as
\begin{eqnarray}\label{Omega-b}
\Omega_\beta=
\begin{cases}
\sqrt{1-\frac{m^2}{\omega^2}},& \omega>m,\\
0,& \omega\leq m.
\end{cases}
\end{eqnarray}
In both the two cases, one can prove that $\Omega\in[0,1]$, with $\Omega=1$ corresponding to the massless case. One can refer to Appendix \ref{various cases} for the derivations of Eqs. \eqref{g} and \eqref{Omega-b}.

{\it 2. $\eta$.}
If the environment is in an equilibrium state which satisfies the  Kubo-Martin-Schwinger (KMS) condition \cite{Kubo1957,Martin1959,Haag1967} (which is true for the models considered in the present paper),  the Wightman function satisfies the following property
\begin{equation}
G(\Delta\tau-i\beta_K)=G(-\Delta\tau),
\end{equation}
where $\beta_K={1}/{T_{K}}$ is a positive parameter called the inverse KMS temperature. Then, we can  obtain that  $\mathcal{G}^+(-\omega)=e^{-\beta_K\omega}\mathcal{G}^+(\omega)$, where $\mathcal{G}^+(\omega)=\int^{\infty}_{-\infty}G^{+}(\Delta\tau) e^{i\omega\Delta \tau} d\Delta \tau$, and define a dimensionless factor $\eta$ as
\begin{eqnarray}\label{etad}
\eta=\frac{D_{+}^{(\alpha\varrho)}+D_{-}^{(\alpha\varrho)}}{D_{+}^{(\alpha\varrho)}-D_{-}^{(\alpha\varrho)}}=\frac{e^{\omega\beta_K}+1}{e^{\omega\beta_K}-1}=\coth{\frac{\omega\beta_K}{2}}.
\end{eqnarray}
The factor $\eta$ can be re-expressed with the downward and upward transition rates $\mathcal{R}^{\downarrow}=\mathcal{R}_{|h\rangle\rightarrow|l\rangle}$, $\mathcal{R}^{\uparrow}=\mathcal{R}_{|l\rangle\rightarrow|h\rangle}$ between
any two energy eigenstate $|h\rangle$ (high), and $|l\rangle$ (low) as
\begin{eqnarray}\label{etad2}
\eta=\frac{\mathcal{R}^{\downarrow}+\mathcal{R}^{\uparrow}}{\mathcal{R}^{\downarrow}-\mathcal{R}^{\uparrow}}=2N_{K}+1
=\coth{{\omega}\/{2T_{K}}},
\end{eqnarray}
where $N_{K}$ is the effective particle number, and the KMS temperature $T_{K}={1}/{\beta_{K}}=\omega\,[\ln(1+1/N_{K})]^{-1}$.
For the acceleration case,
\bea\label{eta-a}
\eta_a=\coth\frac{\pi\omega}{a}=\coth\frac{\omega}{2T_U},
\eea
where $T_U=\frac{a}{2\pi}$ is the Unruh temperature, and  for the  thermal bath case
\bea\label{eta-b}
\eta_\beta=\coth\frac{\beta\omega}{2}=\coth\frac{\omega}{2T}.
\eea
Here, one can see Appendix \ref{various cases} for details. Moreover,
it is obvious that $\eta\in[1,+\infty]$, with $\eta=1$  corresponding to the case when the (Unruh) temperature is zero.

{\it 3. $\lambda$.}
In our model, $\Gamma^{(11)}=\Gamma^{(22)}$ and $\Gamma^{(12)}=\Gamma^{(21)}$. Then we can define a factor $\lambda$ as
\begin{eqnarray}\label{lambdad}
\lambda=\frac{\Gamma^{(\alpha\varrho)}\big|_{\alpha\neq\varrho}}
{\Gamma^{(\alpha\varrho)}\big|_{\alpha=\varrho}},
\end{eqnarray}
Similarly, the factor $\lambda$ can be re-expressed with the upward and downward collective transition rates related to the symmetric state $|S\rangle=\frac{1}{\sqrt{2}}(|10 \rangle+|01 \rangle)$ and anti-symmetric state $|A\rangle=\frac{1}{\sqrt{2}}(|10 \rangle-|01 \rangle)$ as
\begin{eqnarray}\label{lambdad2}
\lambda=\frac{\mathcal{R}^{\downarrow}_S-\mathcal{R}^{\downarrow}_A}{\mathcal{R}^{\downarrow}_S+\mathcal{R}^{\downarrow}_A}
=\frac{\mathcal{R}^{\uparrow}_S-\mathcal{R}^{\uparrow}_A}{\mathcal{R}^{\uparrow}_S+\mathcal{R}^{\uparrow}_A}\;,
\end{eqnarray}
where, the collective rates are respectively $\mathcal{R}^{\downarrow}_S=\mathcal{R}_{|E\rangle\rightarrow |S\rangle}=\mathcal{R}_{|S\rangle\rightarrow |G\rangle}$, $\mathcal{R}^{\downarrow}_A=\mathcal{R}_{|E\rangle\rightarrow |A\rangle}=\mathcal{R}_{|A\rangle\rightarrow |G\rangle}$, $\mathcal{R}^{\uparrow}_S=\mathcal{R}_{|G\rangle\rightarrow |S\rangle}=\mathcal{R}_{|S\rangle\rightarrow |E\rangle}$ and $\mathcal{R}^{\uparrow}_A=\mathcal{R}_{|G\rangle\rightarrow |A\rangle}=\mathcal{R}_{|A\rangle\rightarrow |E\rangle}$, with $|E\rangle=|11\rangle$ and $|G\rangle=|00\rangle$. From Eq. \eqref{lambdad2} , it is straightforward to show  that $|\lambda|\leq1$.
For the acceleration case, $\lambda_a$ can be obtained as
\begin{eqnarray}\label{lambda a}
\lambda_a=\frac{4a^2}{m^2}\frac{\int_{m\/a}^{\infty}{1\/{a L}}{\sin\big(a L\sqrt{x^2-{m^2/{a^2}}}\big)}\;K_{i2\omega/a}\left(2x\right) dx }{K_{1+i\omega/a}\left(m/a\right)K_{-1+i\omega/a}\left(m/a\right)
-K_{i\omega/a}^2\left(m/a\right)},
\end{eqnarray}
which can be written as a function of dimensionless variables, i.e.,  $\lambda_a(m/\omega,a/\omega,L\omega)$. Here, $L$ is the separation between the two detectors. For the thermal bath case,
\begin{eqnarray}\label{lambda b}
\lambda_\beta=\frac{\sin(L\omega\Omega_\beta)}{L\omega\Omega_\beta}.
\end{eqnarray}
Recall that $\Omega_\beta$ has been given in Eq. (\ref{Omega-b}).
The derivations  of Eqs. \eqref{lambda a} and \eqref{lambda b} are given in  Appendix \ref{various cases}, and the details of the properties of $\lambda_a$ are referred to  Appendix \ref{the properties lambda}.

{\it 4. $\zeta^{+}$ and $\zeta^{-}$.}
We define two factors
\bea\label{zeta+gamma d}
\zeta^{+}=\frac{D_{0}^{(11)}+D_{0}^{(12)}}{D_{+}^{(11)}+D_{-}^{(11)}},\;\;\zeta^{-}=\frac{D_{0}^{(11)}-D_{0}^{(12)}}{D_{+}^{(11)}+D_{-}^{(11)}}.
\eea
With the help of Eq. \eqref{D2} and the parameters defined previously, $\zeta^{\pm}$ can be re-expressed as
\bea\label{zeta+gamma re}
&&\zeta^{\pm}=\frac{{\cal G}^{(11)}(0)\pm{\cal G}^{(12)}(0)}{2\left[{\cal G}^{(11)}(\omega)+{\cal G}^{(11)}(-\omega)\right]}=\frac{2T_{K}}{\omega}\frac{1}{\eta}\frac{\Omega(0)}{\Omega(\omega)}\frac{1\pm\lambda(0)}{2}.
\eea
It is obvious that $\zeta^{\pm}\geq0$. Now, for the acceleration case, substituting Eqs. \eqref{g}, \eqref{eta-a} and \eqref{lambda a} into Eq. \eqref{zeta+gamma re}, one can obtain
\begin{eqnarray}\label{gammaa}
\zeta^{\pm}_a=\frac{{m^2}\left[K_{1}^2\left(m/a\right)-K_{0}^2\left(m/a\right)\right]\pm 4{a^2}\int_{m\/a}^{\infty}{\frac{\sin\big(a L\sqrt{x^2-{m^2/{a^2}}}\big)}{a L}}\;K_{0}\left(2x\right) dx }{2{m^2}\left[K_{1+i\omega/a}\left(m/a\right)K_{-1+i\omega/a}\left(m/a\right)
-K_{i\omega/a}^2\left(m/a\right)\right]\cosh\left(\pi\omega/a\right)},
\end{eqnarray}
which can be written as a function of dimensionless variables, i.e., $\zeta^{\pm}_a(m/\omega,a/\omega,L\omega)$ (See Appendix \ref{the properties zeta} for the details of the properties of $\zeta^{\pm}_a$). Similarly, for the thermal bath case,
\begin{eqnarray}\label{zetab+gammab}
\zeta^{+}_{\beta}=
\begin{cases}
\frac{\tanh(\omega\beta/2)}{\omega\beta/2},& m=0,\\
0,& m\neq0,
\end{cases}\quad\quad\zeta^{-}_{\beta}=0.
\end{eqnarray}

With the  parameters defined above,  $D_{\kappa}^{(\alpha\varrho)}$ in Eq. \eqref{D2} can  be re-expressed as
\begin{eqnarray}\label{Dij2}
&&D_{\pm}^{(11)}=D_{\pm}^{(22)}=\frac{1}{2}\left(\eta\pm1\right)\Omega\Gamma_{0},\;\;\;\;\;D_{\pm}^{(12)}=D_{\pm}^{(21)}=\frac{1}{2}\left(\eta\pm1\right)\lambda\;\Omega\Gamma_{0}\nonumber\\
&&D_{0}^{(11)}=D_{0}^{(22)}=\frac{1}{2}(\zeta^{+}+\zeta^{-})\eta\Omega\Gamma_{0},\;\;\;\;\;D_{0}^{(12)}=D_{0}^{(21)}=\frac{1}{2}(\zeta^{+}-\zeta^{-})\eta\Omega\Gamma_{0}.
\end{eqnarray}
Now, we choose to work in the coupled basis $\{|E \rangle,|S \rangle,|A \rangle,|G \rangle\}$,
where $|E \rangle=|11 \rangle$, $|S \rangle=\frac{1}{\sqrt{2}}(|10 \rangle+|01 \rangle)$, $|A \rangle=\frac{1}{\sqrt{2}}(|10 \rangle-|01 \rangle)$ and $|G \rangle=|00 \rangle$. For simplicity, we assume that the initial density matrix is in the X form, i.e. the only nonzero elements are those along the diagonal and anti-diagonal of the density matrix in the coupled basis $\{|E \rangle,|S \rangle,|A \rangle,|G \rangle\}$, then the X form will be maintained during evolution \cite{xstate}. Thus, a set of equations which describe the time evolution of the density matrix elements can be expressed as
\begin{eqnarray}\label{evolution}
&&\dot{{\bf X}}(\tau)=-{\bf U}(\eta,\lambda,\zeta^{-})\;{\bf X}(\tau),\;\;\;\;\;\dot{\rho}_{EG}(\tau)=-(1+4\zeta^{+})\rho_{EG}(\tau),\nonumber\\%\;\;\;\;\;\;\;\;\;\dot{\rho}_{GE}(\tau)=-\rho_{GE}(\tau)
&&\dot{\rho}_{SA}(\tau)=-(1+2\zeta^{-})\rho_{SA}(\tau)+2\zeta^{-}\rho_{AS}(\tau),%\;\;\dot{\rho}_{AS}(\tau)=-(1+\zeta^{-})\rho_{AS}(\tau)+\zeta^{-}\rho_{SA}(\tau).\hspace{0.5cm}
\end{eqnarray}
where $\rho_{IJ}(\tau)=\langle I|\rho(\tau)|J\rangle$, $I,J\in\{G,E,A,S\}$,
and $\dot{\rho}_{IJ}(\tau)={d \rho_{IJ}(\tau)}/{d r}$ is the derivative with respect to the dimensionless time $r=\eta\Omega\Gamma_0\tau$. The column vector ${\bf X}(\tau)$
is defined as ${\bf X}(\tau)=\Big(\rho_{E}(\tau)\;\;\rho_{S}(\tau)\;\;\rho_{A}(\tau)\;\;\rho_{G}(\tau)\Big)^{T}$.
Here, for brevity, we have abbreviated the diagonal terms of the density matrix elements as $\rho_{I}(\tau)=\rho_{II}(\tau)$. The coefficient  matrix ${\bf U}(\eta,\lambda,\zeta^{-})$ can be expressed as
\begin{eqnarray}\label{U}
{\bf U}(\eta,\lambda,\zeta^{-})=
\begin{pmatrix} \frac{\eta+1}{\eta}&-\frac{(\eta-1)(1+\lambda)}{2\eta}&-\frac{(\eta-1)(1-\lambda)}{2\eta}&0\\
                -\frac{(\eta+1)(1+\lambda)}{2\eta}&1+\lambda+2\zeta^{-}&-2\zeta^{-}&-\frac{(\eta-1)(1+\lambda)}{2\eta}\\
                -\frac{(\eta+1)(1-\lambda)}{2\eta}&-2\zeta^{-}&1-\lambda+2\zeta^{-}&-\frac{(\eta-1)(1-\lambda)}{2\eta}\\
                0&-\frac{(\eta+1)(1+\lambda)}{2\eta}&-\frac{(\eta+1)(1-\lambda)}{2\eta}&\frac{\eta-1}{\eta}
\end{pmatrix},
\end{eqnarray}
Since ${\rho}_{G}+{\rho}_{E}+{\rho}_{A}+{\rho}_{S}=1$, only three of the first four equations in Eq. \eqref{evolution} are independent.
The general solution of Eq. \eqref{evolution}  can  be written in the following form
\bea\label{general solution}
&{\bf X}(\tau)={\bf M}_0(\eta)+\sum\limits_{i=1}^{3}{\bf M}_{i}\big(\rho(0),\eta,\lambda,\zeta^{-}\big)\;[\Theta(\tau)]^{\xi_{i}},\nonumber\\
&\rho_{SA}(\tau)=\frac{\rho_{SA}(0)+\rho_{SA}(0)}{2}\;\Theta(\tau)+\frac{\rho_{SA}(0)-\rho_{SA}(0)}{2}\;[\Theta(\tau)]^{1+4\zeta^{-}},\nonumber\\
&\rho_{EG}(\tau)=\rho_{EG}(0)\;[\Theta(\tau)]^{1+4\zeta^{+}},
\eea
where
\bea\label{Theta}
\Theta(\tau)=e^{-\eta\Omega\Gamma_0\tau},
\eea
which ranges from 0 to 1, and  it determines the entanglement evolution with time.
Here,  $\xi_{i}(\eta,\lambda,\zeta^{-})$ and ${\bf M}_{i}\big(\rho(0),\eta,\lambda,\zeta^{-}\big)$  are respectively the eigenvalues and corresponding eigenvectors of the coefficient square matrix ${\bf U}(\eta,\lambda,\zeta^{-})$. The eigenvalues $\xi_{i}$ satisfy the following equation,
\bea\label{ch-eq}
\xi[\xi^3-4(1+\zeta^{-})\xi^2+(5+12\zeta^{-}+\lambda^2\eta^{-2}-2\lambda^2)\xi-2(1+4\zeta^{-}-\lambda^2)]=0.
\eea
Obviously, there is a definite zero root which we label as $\xi_0=0$, and the corresponding eigenvector ${\bf M}_{0}(\eta)$ can be obtained as
\bea\label{final-rho}
{\bf M}_{0}(\eta)=\Bigg(\frac{(\eta+1)^2}{4\eta^2}\;\;\; \frac{\eta^2-1}{4\eta^2}\;\;\; \frac{\eta^2-1}{4\eta^2}\;\;\; \frac{(\eta-1)^2}{4\eta^2}\Bigg)^{T}.
\eea
The other three non-negative roots are labeled as $\xi_1,\;\xi_2,\;\xi_3$. Hereafter, the subscript ``${i}$'' of $\xi_{i}$ runs from $1$ to $3$ except otherwise stated. Note that the explicit expressions for $\xi_{i}(\eta,\lambda,\zeta^{-})$ and ${\bf M}_{i}\big(\rho(0),\eta,\lambda,\zeta^{-}\big)$  which are not directly mentioned in the following  discussions are not shown here.

We characterize the degree of entanglement by concurrence \cite{W. K. Wootters}, which ranges from 0 for separable states, to 1 for maximally entangled states. For the X-type states, the concurrence takes the following form \cite{R. Tanas}
\begin{eqnarray}\label{C}
C[\rho(\tau)]=\mathrm{max}\{0,K_{1}(\tau),K_{2}(\tau)\},
\end{eqnarray}
where
\begin{eqnarray}\label{k1}
&&K_{1}(\tau)=\sqrt{[\rho_{A}(\tau)-\rho_{S}(\tau)]^{2}-[\rho_{AS}(\tau)-\rho_{SA}(\tau)]^{2}}-2\sqrt{\rho_{G}(\tau)\rho_{E}(\tau)},\nonumber\\
&&K_{2}(\tau)=2|\rho_{GE}(\tau)|-\sqrt{[\rho_{A}(\tau)+\rho_{S}(\tau)]^{2}-[\rho_{AS}(\tau)+\rho_{SA}(\tau)]^{2}}.
\end{eqnarray}
Substituting  Eq. \eqref{general solution} into  Eq. \eqref{C}, one can  obtain the concurrence $C$, which can be formally written as $C[\Omega\tau,\rho(0),\eta,\lambda,\zeta^{\pm}]$.

\section{Entanglement dynamics of a uniformly accelerated quantum system coupled with massive scalar fields}
In this section, we study the entanglement dynamics of a quantum system composed of two uniformly accelerated Unruh-DeWitt detectors coupled with massive scalar fields in the Minkowski vacuum. With the help of the  parameters $\{\Omega,\eta,\lambda,\zeta^{\pm}\}$ defined in the preceding section, we study the time-delay effect (related to the parameter $\Omega$), entanglement degradation effect (related to the parameters $\eta$, $\zeta^{\pm}$), and entanglement generation effect (related to the parameter $\lambda$) in entanglement evolution. We also discuss the Unruh and anti-Unruh effects by comparing the entanglement generated in the acceleration case with that in the thermal bath case.

\subsection{The time-delay effect}
From the time evolution of the density matrix Eqs. \eqref{general solution} and \eqref{Theta}, it is obvious that the factor $\Omega$ plays the role of delaying the entanglement evolution, since its range is $ 0\leq\Omega \leq1 $.
This time-delay effect of the entanglement evolution is general, which is independent of the initial state of the system,  or the feature of entanglement evolution (e.g. entanglement generation, degradation, revival, etc). That is, the evolution time for the quantum system coupled with massive fields is always $\Omega^{-1}$  times that of the massless case. This time-delay effect is advantageous to entanglement protection, but disadvantageous to  entanglement generation.

For uniformly accelerated Unruh-DeWitt detectors coupled to massive scalar fields, the factor $\Omega_a$ can be written as a function of $m/\omega$ and $a/\omega$, whose analytical expression and the numerical values are shown in Eq. (\ref{g}) and Fig. \ref{Omega(m,a)} respectively.
In Ref. \cite{Y. B. Zhou}, the properties of  $\Omega_a\big(m/\omega,a/\omega\big)$ have been studied in detail. The properties of  $\Omega_a$ under some limiting conditions, such as the low and high acceleration limit, and the low and high mass limit,  are shown in Appendix \ref{the properties Omega}. In the following, we analyze the influences of the mass of the field and the acceleration on the time-delay effect in entanglement evolution.
\begin{figure}[!htbp]
\centering
\includegraphics[width=0.49\textwidth]{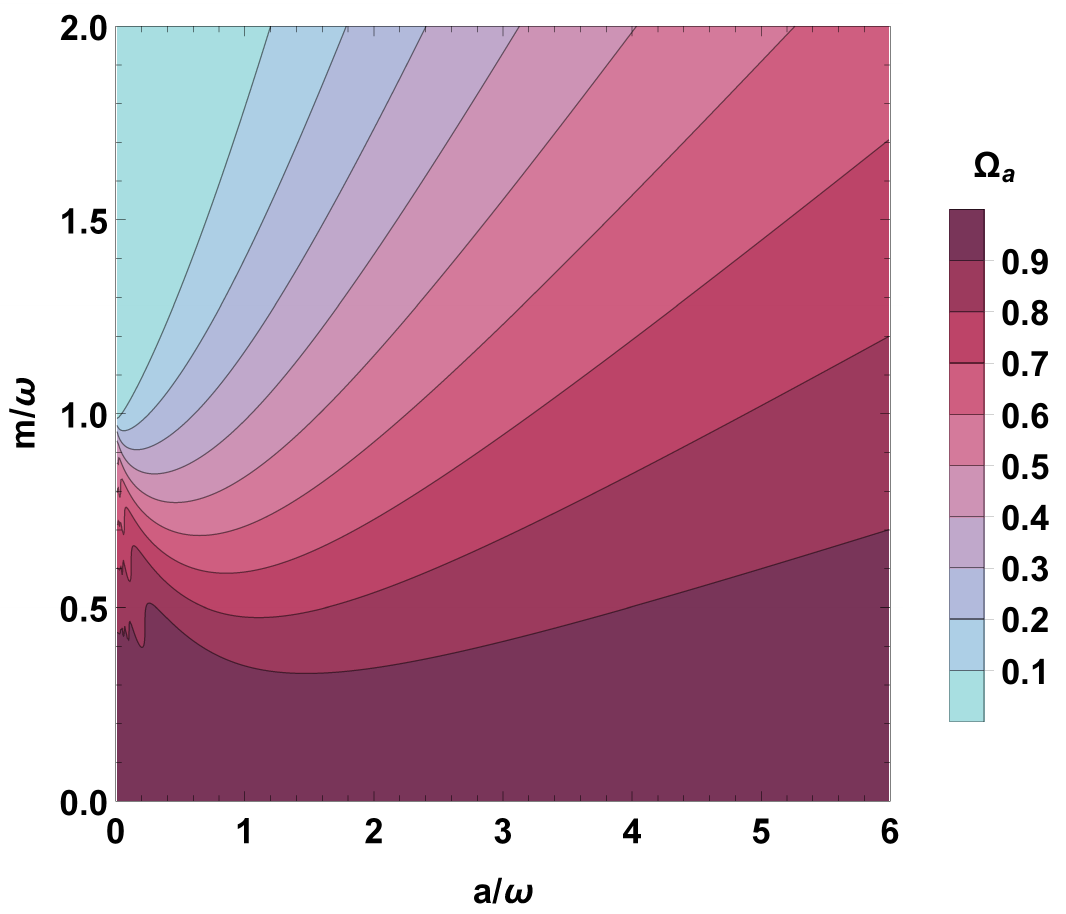}
\caption{\label{Omega(m,a)}
 The contour map of the factor $\Omega_a$ in parameter space $(a/\omega, m/\omega)$. }
\end{figure}

\subsubsection{The mass effects}
\begin{figure}[!htbp]
\centering
\includegraphics[width=0.49\textwidth]{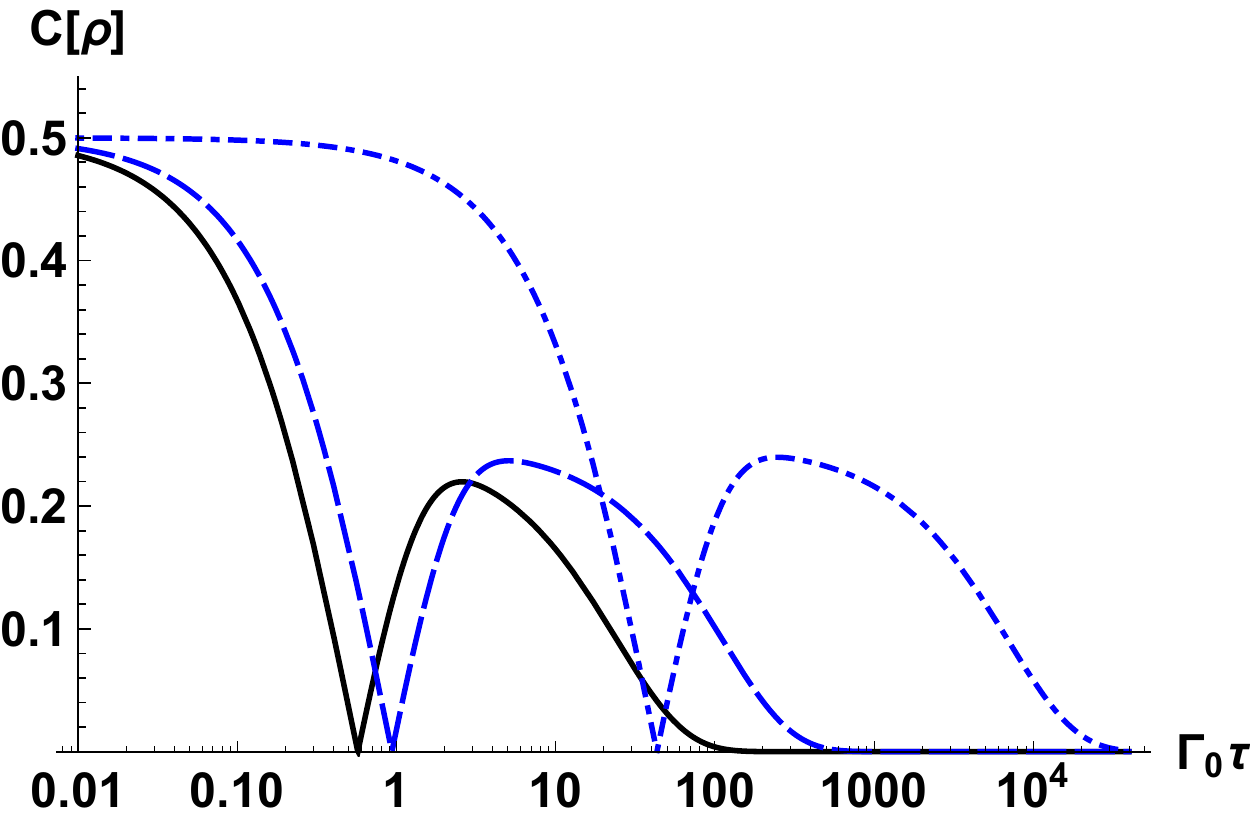}
\includegraphics[width=0.5\textwidth]{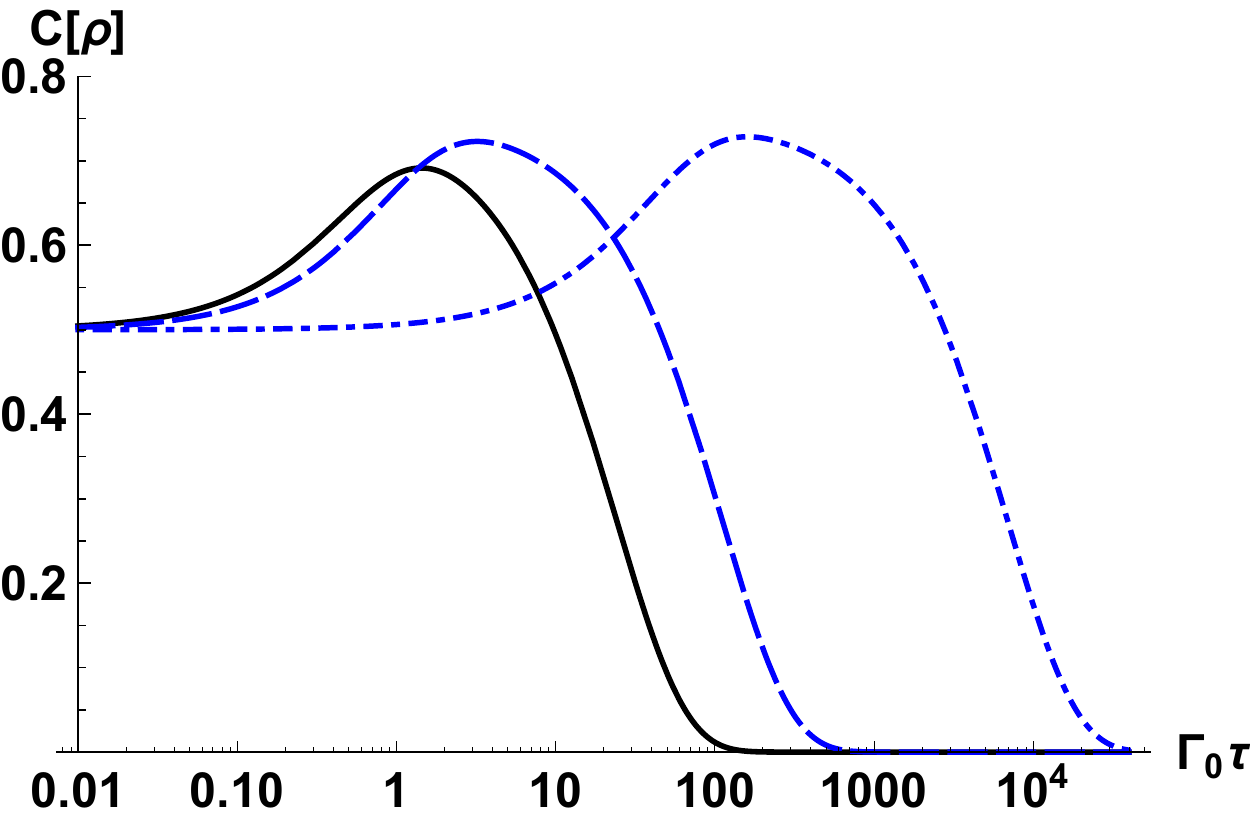}
\caption{\label{ger1}
Comparison between the dynamics of concurrence for uniformly accelerated quantum systems coupled with massive scalar fields (blue lines) and massless scalar fields (black lines) initially prepared in $\frac{1}{2}|A\rangle+\frac{\sqrt{3}}{2}|S\rangle$ (left) and $\frac{\sqrt{3}}{2}|A\rangle+\frac{1}{2}|S\rangle$ (right), with $L\omega=0.5$, $a/\omega=0.1$. The black solid, blue dashed and dot-dashed lines correspond to $m/\omega=0,\;0.8,\;1.2$ respectively.}
\end{figure}

In this part, we  focus on the effects of mass on the time-delay effect. Compared with the massless case, the entanglement evolution in the massive case is generally slower since $\Omega_a\leq1$.
When  $m\to 0$, $\Omega_a$  tends to that in the massless case, i.e.,  $\Omega_{0a}=1$. Also, it can be proved that $d\Omega_a/dm\leq0$ (See Eq. \eqref{app-c2} in Appendix \ref{the properties Omega}), so the larger the mass $m$, the slower the evolution. Especially, when $m/\omega\geq1$, the time delay  will increase  exponentially.
(See Appendix \ref{the properties Omega} for the details of the properties of $\Omega_a$).
This time-delay phenomenon is shown in Fig. \ref{ger1}.

\subsubsection{The acceleration effects}
Comparing $\Omega_a$ and $\Omega_\beta$ given in Eqs. (\ref{g}) and \eqref{Omega-b}, one finds that, for uniformly accelerated Unruh-DeWitt detectors in the Minkowski vacuum, $\Omega_a$ is not only related to the field mass $m/\omega$, but also to the acceleration $a/\omega$. However, for the static ones in a thermal bath, $\Omega_\beta$ is only related to $m/\omega$ but not to the temperature of the bath $T_U$  \cite{Zhou2020}. This leads to  significant differences between the two cases.

\begin{figure}[!htbp]
\centering
\includegraphics[width=0.5\textwidth]{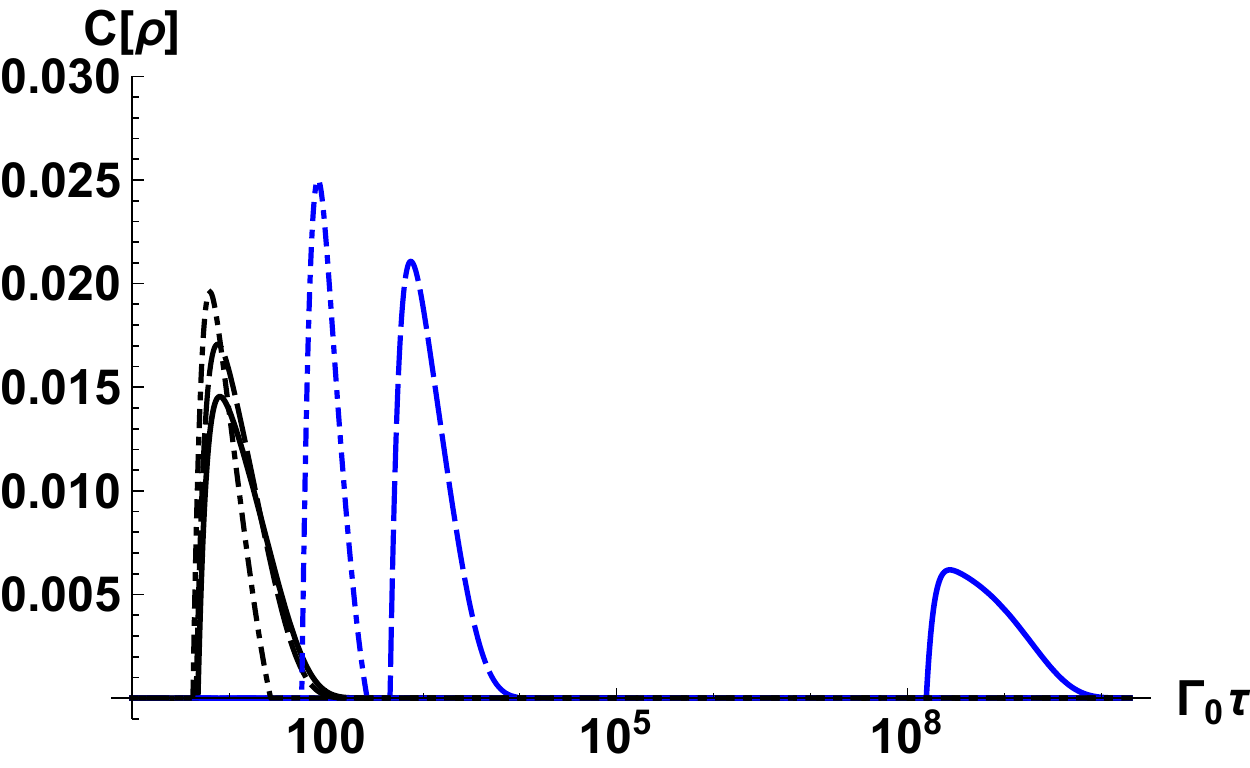}
\includegraphics[width=0.49\textwidth]{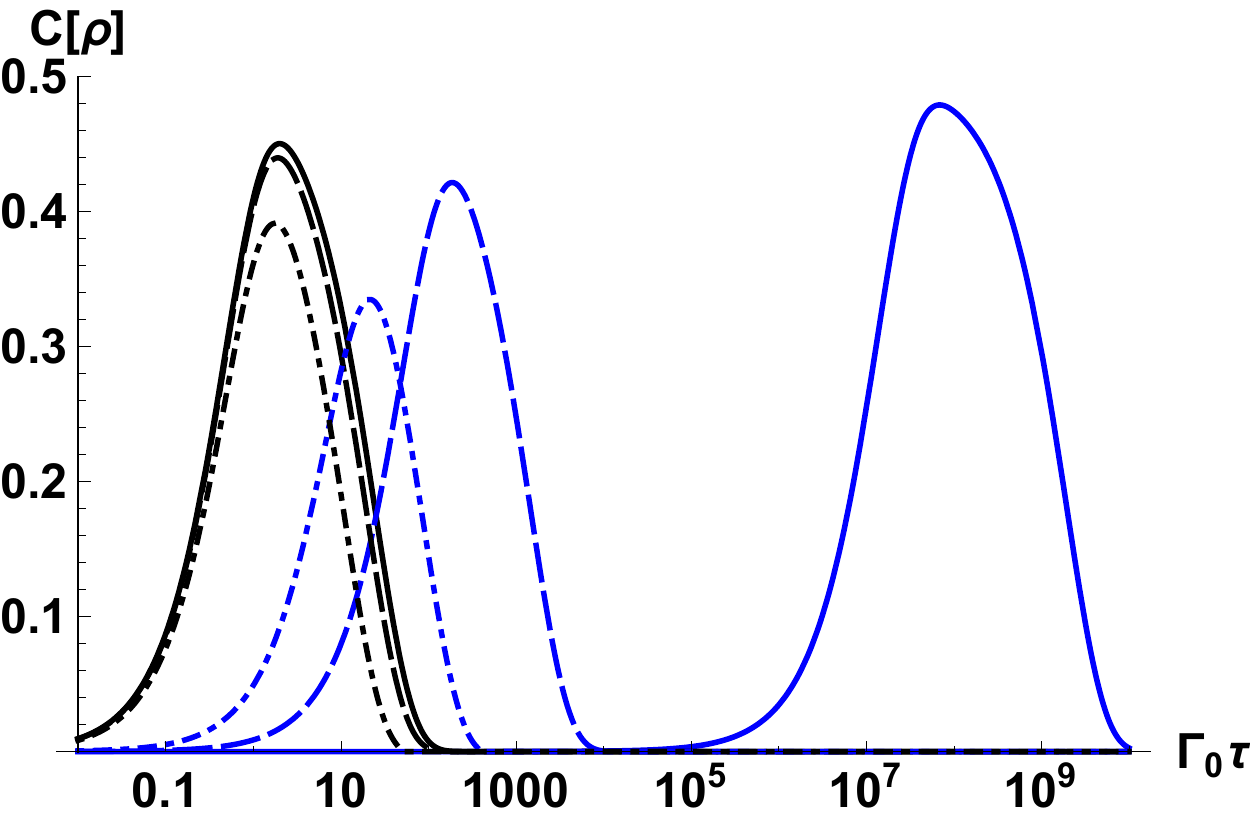}
\caption{\label{gec2}
Comparison between the dynamics of concurrence for uniformly accelerated quantum systems coupled with massive scalar fields (blue lines, $m/\omega=2$) and massless scalar fields (black lines,  $m/\omega=0$) initially prepared in $|E\rangle$ (left) and $\frac{\sqrt{2}}{2}|A\rangle\pm\frac{\sqrt{2}}{2}|S\rangle$ (i.e. $|10\rangle$ or $|01\rangle$) (right), with $L\omega=0.5$. The solid, dashed and dot-dashed lines  correspond to $a/\omega$=0.1, 0.5, 1 respectively.}
\end{figure}

Firstly, as shown in Fig. \ref{Omega(m,a)}, when $m<\omega$, the factor $\Omega_a$ oscillates as the acceleration increases when the acceleration is small, and increases monotonically with the acceleration when the acceleration is large enough. When  $m\geq\omega$, the factor $\Omega_a$  increases monotonically with the acceleration. In either case, ${\Omega_{a}}\to1$ when the acceleration tends to infinity (See Appendix \ref{the properties Omega} for the proof). That is, the time-delay effect brought about by field being massive can be counteracted by a large enough acceleration (See also Fig.~\ref{gec2}).
This is in sharp contrast to the case of a static quantum system in a thermal bath, where this time delay brought about by the field being massive is not affected by the temperature.
Secondly, in the thermal case, when $m/\omega\geq1$, $\Omega_\beta=0$, which means that the quantum system is locked up in its initial state $\rho(\tau)=\rho(0)$, and the concurrence is a constant \cite{Zhou2020}. However, in the acceleration case, as long as $a\neq0$ and $m\neq\infty$, then $\Omega_a\neq0$, and so the entanglement  still evolves, although the evolution may be very slow because $\Omega_a\to 0$ when $a\to 0$ or $m\to\infty$.

\subsection{The entanglement degradation effect}
The entanglement degradation is caused by the dissipation and decoherence due to the quantum system with the environment, which will be discussed in details as follows.

\subsubsection{Entanglement degradation caused by dissipation}

Now, we discuss the entanglement degradation effect  caused by dissipation characterized by the factor $\eta$ defined in Eq. \eqref{etad}, which can be written as $\eta= \coth{\left[\omega/{(2T_{K})}\right]}$, where $T_{K}$ is the Unruh temperature in the acceleration case and  the environment temperature in the thermal case.

Firstly, the larger the factor $ \eta $, the smaller the asymptotic entanglement of the quantum system.
When  the separation between the two detectors is nonvanishing, $\lambda<1$, and it can be found from Eq. \eqref{ch-eq} that the eigenvalues $\xi_{i}$ of the coefficient matrix ${\bf U}$ defined in Eq. \eqref{U} are positive. Then, from Eq. \eqref{general solution}, one can  obtain the asymptotic state as
\bea
\rho_{E}(\infty)=\frac{(\eta-1)^2}{4\eta^2},\;\;\;\rho_{S}(\infty)=\rho_{A}(\infty)=\frac{\eta^2-1}{4\eta^2},
\;\;\;\rho_{G}(\infty)=\frac{(\eta+1)^2}{4\eta^2},
\eea
and $\rho_{IJ}(\infty)=0 $ ($I\neq J$). Plugging the asymptotic state into Eq. \eqref{C}, we can get
\bea\label{Asymptotic entanglement1}
K_1 (\infty) = K_2 (\infty) =\frac{1}{2}\left(\eta^{-2}-1\right)\leq0,
\eea
so $ C (\infty) = 0 $, which indicates that the final state is a separable state.
When the separation between the two detectors is vanishing, we have $\lambda=1$ and $\zeta^{-}=0$, and the asymptotic state can be obtained as
\bea\label{as-state-gamma0-lambda1}
&&\rho_{E}(\infty)=\frac{(\eta-1)^2}{1+3\eta^2}\big[1-\rho_{A}(0)\big],\;\;\;\rho_{S}(\infty)=\frac{\eta^2-1}{1+3\eta^2}\big[1-\rho_{A}(0)\big],\nonumber\\
&&\rho_{G}(\infty)=\frac{(\eta+1)^2}{1+3\eta^2}\big[1-\rho_{A}(0)\big],\;\;\;\rho_{A}(\infty)=\rho_{A}(0),
\eea
and $\rho_{IJ}(\infty)=0 $ ($I\neq J$). Similarly, we can get
\bea\label{Asymptotic entanglement2}
&&K_{1}(\infty)=\frac{\big|4\eta^2\rho_{A}(0)+1-\eta^2\big|-2(\eta^2-1)\big(1-\rho_{A}(0)\big)}{3\eta^2+1},\nonumber\\
&&K_{2}(\infty)=-\frac{2(\eta^2+1)\rho_{A}(0)+\eta^2-1}{3\eta^2+1}\leq0.
\eea
In this case, if $\rho_{A}(0)>\frac{3(\eta^2-1)}{2(3\eta^2-1)}$, then,
\bea\label{Asymptotic entanglement3}
C(\infty)=K_{1}(\infty)=\frac{4[1-\rho_{A}(0)]}{3\eta^2+1}+2\rho_{A}(0)-1>0,\hspace{0.6cm}
\eea
which indicates that there exists asymptotic entanglement related to the initial state.
It is found that the mass of the field cannot affect the asymptotic entanglement, so the result is the same as that in the massless case \cite{Benatti}. However, this is different from the  result when the qubits are accelerating in an environment with nonzero background temperature
 \cite{M Lima2020}.
From Eqs. \eqref{Asymptotic entanglement1} and \eqref{Asymptotic entanglement2}, it is easy to prove that $\frac{d K_{1}(\infty)}{d\eta}\leq0,\;\frac{d K_{2}(\infty)}{d\eta}\leq0$, which indicates that  $\eta$ plays the  role of entanglement degradation.

Secondly, $ \eta $  appears in  $ \Theta(\tau)=e ^ {-\eta\Omega \Gamma_0 \tau} $, which shows that it accelerates the evolution of the system. In other words, it speeds up the disentanglement of a quantum system when there is a finite separation between the two detectors.
\begin{figure}[!htbp]
\centering
\includegraphics[width=0.49\textwidth]{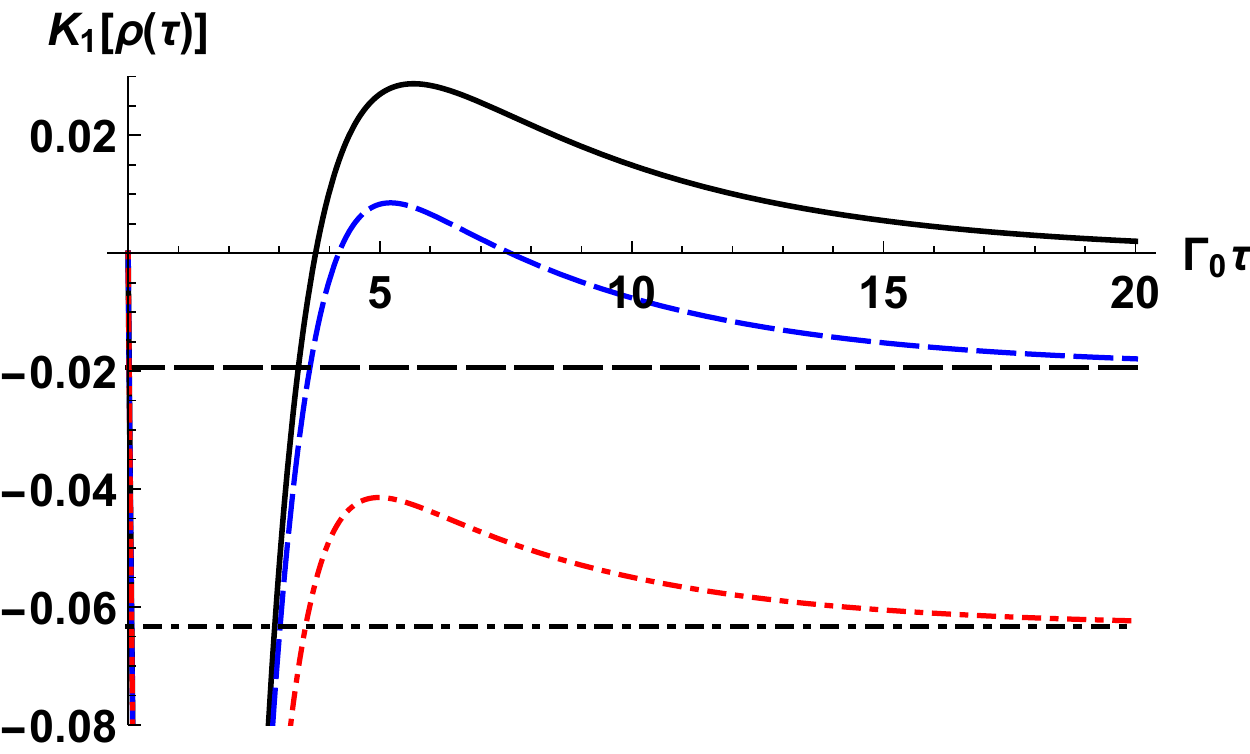}
\includegraphics[width=0.5\textwidth]{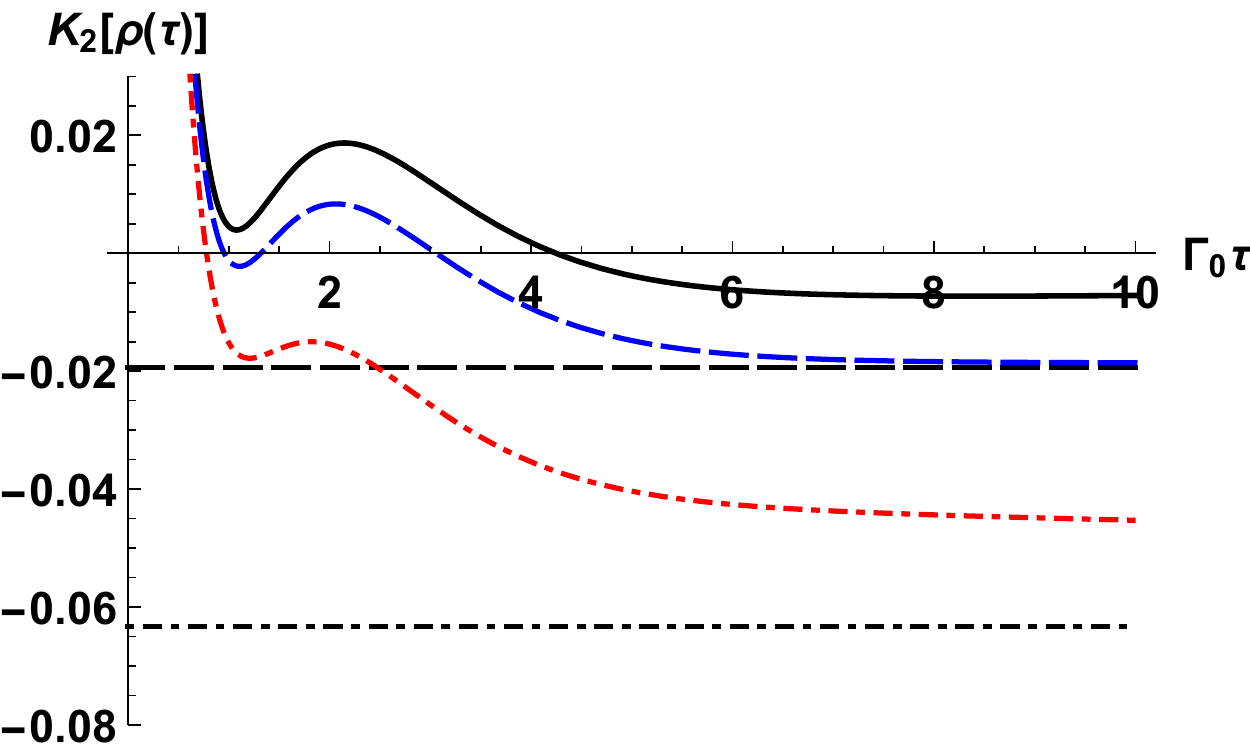}
\caption{\label{Dissipation}
The time evolution of the concurrence parameters $K_1$ (left, with  $\lambda=0.8$, $\zeta^{-}=0$) and $K_2$ (right, with  $\lambda=0.98$, $\zeta^{+}=0$) with different dissipation factor $\eta=\coth{\left[\omega/(2T_{K})\right]}$, for two-atom system prepared in $|E\rangle$ (left), $\sqrt{0.13}|G\rangle+\sqrt{0.87}|E\rangle$ (right).  The black solid (horizontal axis), blue (black) dashed and red (black) dot-dashed lines are the evolution-lines (asymptotes) corresponding to $\eta=1,\;1.02,\;1.07$ respectively.}
\end{figure}

Thirdly, $ \eta $ also affects the value of the eigenvector ${\bf M}_{i}\big(\rho(0),\eta,\lambda,\zeta^{-}\big)$ (i.e., the amplitude functions in Eq. \eqref{general solution}) and the eigenvalues $\xi_{i}(\eta,\lambda,\zeta^{-})$ of the coefficient matrix ${\bf U}(\eta,\lambda,\zeta^{-})$, which affect  the phenomena of entanglement evolution,  such as entanglement creation, revival,  enhancement, etc. We show the effect of $\eta$ on  the concurrence coefficients $K_{1}$ and $K_{2}$ defined in Eq. \eqref{k1} in Fig. \ref{Dissipation}.  There one can see that the larger the $ \eta $ (the temperature $T_{K}$),  the more the curve of  $K_{1(2)}$ moves downwards, and the smaller the concurrence is. So the phenomena of entanglement generation, enhancement, and revival will not show up when the factor $ \eta $ becomes large enough.

To summarize, the factor $\eta$, which describes the thermal dissipative effect of the environment, plays the role of degrading entanglement in entanglement dynamics. It is only related to the Unruh temperature $T_{U}=\frac{a}{2\pi}$ in the acceleration case and the environment temperature $T$ in the thermal case. This indicates that, from the aspect of entanglement degradation, an accelerated quantum system in the Minkowski vacuum suffers  the same dissipative effect as that of a static one in a thermal bath at the Unruh temperature,  as expected due to the Unruh effect.

\subsubsection{Entanglement degradation caused by decoherence}

From the evolution equations  of quantum systems \eqref{evolution}, it can be found that the factors $\zeta^{\pm}$ defined in Eq. \eqref{zeta+gamma d} play the role of decoherence. We show that  the decoherence factors cause an additional entanglement degradation in the following.

With the help of the evolution equations of density matrix elements \eqref{evolution}, the first derivative of  $K_{1(2)}(\tau)$  versus time $\tau$ can be written as
\bea\label{derivatives of K12}
&&\frac{d K_1(\tau)}{d\tau}=\left[-\eta K_1(\tau)-4 \zeta^{-}  \eta  K_{11}(\tau)-f_{1}(\tau,\eta)+\lambda\,\big[\rho_{{A}}(\tau)-\rho_{{S}}(\tau)\big]\,h_{1}(\tau,\eta)\right]\Omega\Gamma_0\,, \nonumber\\
&&\frac{d K_2(\tau)}{d\tau}=\left[-\eta K_2(\tau)-4 \zeta^{+}  \eta  K_{21}(\tau)-f_{2}(\tau,\eta)+\lambda\,\big[\rho_{{S}}(\tau)-\rho_{{A}}(\tau)\big]\,h_{2}(\tau,\eta)\right]\Omega\Gamma_0\,,
\eea
where the functions $f_{i}(\tau,\eta)$ and $h_{i}(\tau,\eta)$ can be expressed as
\bea
&&f_{1}(\tau,\eta)=\frac{[(\eta+1)\,\rho_{E}(\tau)+(\eta-1)\,\rho_{G}(\tau)][\rho_{A}(\tau)+\rho_{S}(\tau)]}{K_{12}(\tau)},\nonumber\\
&&f_{2}(\tau,\eta)=\frac{[(\eta+1)\,\rho_{E}(\tau)+(\eta-1)\,\rho_{G}(\tau)][\rho_{A}(\tau)+\rho_{S}(\tau)]}{K_{22}(\tau)}.
\eea
and
\bea
&&h_{1}(\tau,\eta)=\frac{\eta \left[\rho_{{A}}(\tau)+\rho_{{S}}(\tau)\right]}{K_{11}(\tau)}+\frac{K_{1}(\tau)[(\eta+1)\rho_{E}(\tau)+(\eta-1)\rho_{G}(\tau)]}{K_{11}(\tau)K_{12}(\tau)},\nonumber\\
&&h_{2}(\tau,\eta)=\frac{\eta[\rho_{{A}}(\tau)+\rho_{{S}}(\tau)]}{K_{22}(\tau)},
\eea
with
\bea\label{K11-K22}
&K_{11}(\tau)=\sqrt{[\rho_{{A}}(\tau)-\rho_{{S}}(\tau)]^2-[\rho_{{A}{S}}(\tau)-\rho_{{S}{A}}(\tau)]^2},\;\;\;\;K_{12}(\tau)=2\sqrt{\rho_{E}(\tau)\rho_{G}(\tau)},\nonumber\\
&K_{21}(\tau)=2\sqrt{\rho_{EG}(\tau)\rho_{GE}(\tau)},\;\;\;\;K_{22}(\tau)=\sqrt{[\rho_{{A}}(\tau)+\rho_{{S}}(\tau)]^2-[\rho_{{A}{S}}(\tau)+\rho_{{S}{A}}(\tau)]^2}.
\eea
It can be found from Eq. \eqref{derivatives of K12} that, the decoherence factors $\zeta^{-}$ and $\zeta^{+}$ contribute  an additional negative rate of change of $K_1(\tau)$ and $K_2(\tau)$ respectively, which means that the decoherence factors $\zeta^{\pm}$ bring an  additional entanglement degradation. Moreover, from Eq. \eqref{derivatives of K12}, one can see that the larger the factors $\zeta^{\pm}$, the stronger the entanglement degradation. Unlike the dissipation factor $\eta$, which is only related to the acceleration for accelerated detectors coupled with massive scalar field, the decoherence factors  $ \zeta^{\pm}_a(m/\omega,a/\omega,L\omega) $ are related to the acceleration $a$, the mass of the field $m$, and the separation between the detectors $L$. In the following, we focus on the effects of acceleration  and mass  on the entanglement degradation brought about by the decoherence factors $\zeta^{\pm}$ in details.

\paragraph{The acceleration effects}

The decoherence factors $\zeta^{\pm}_{\beta}$ are always zero for a two-detector system coupled with massive scalar field in a thermal bath (see Eq. \eqref{zetab+gammab}), while they are nonzero in the acceleration case (see Eq. \eqref{gammaa}). This indicates that the accelerated detectors are subjected to an additional disentangling effect brought about by the decoherence factors $\zeta^{\pm}$ compared with those immersed in a thermal bath.
\begin{figure}[!htbp]
\centering
\includegraphics[width=0.495\textwidth]{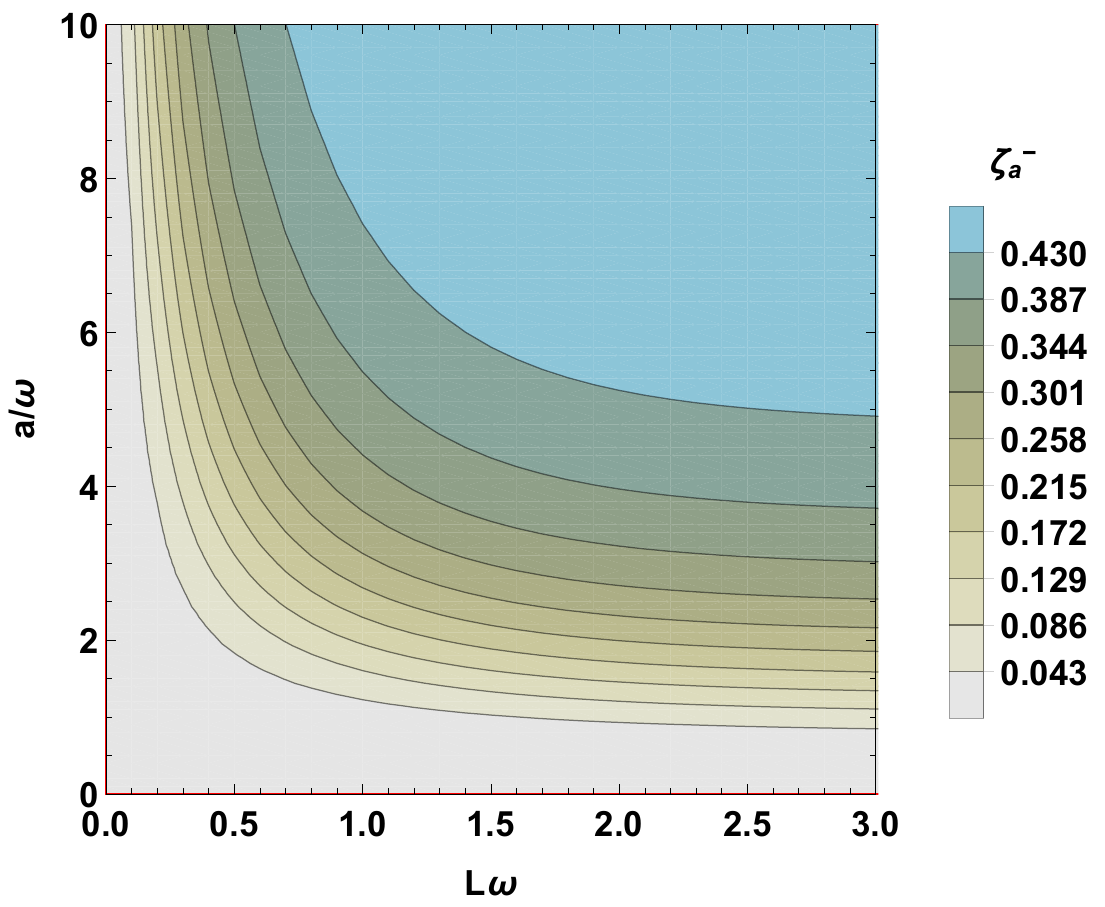}
\includegraphics[width=0.495\textwidth]{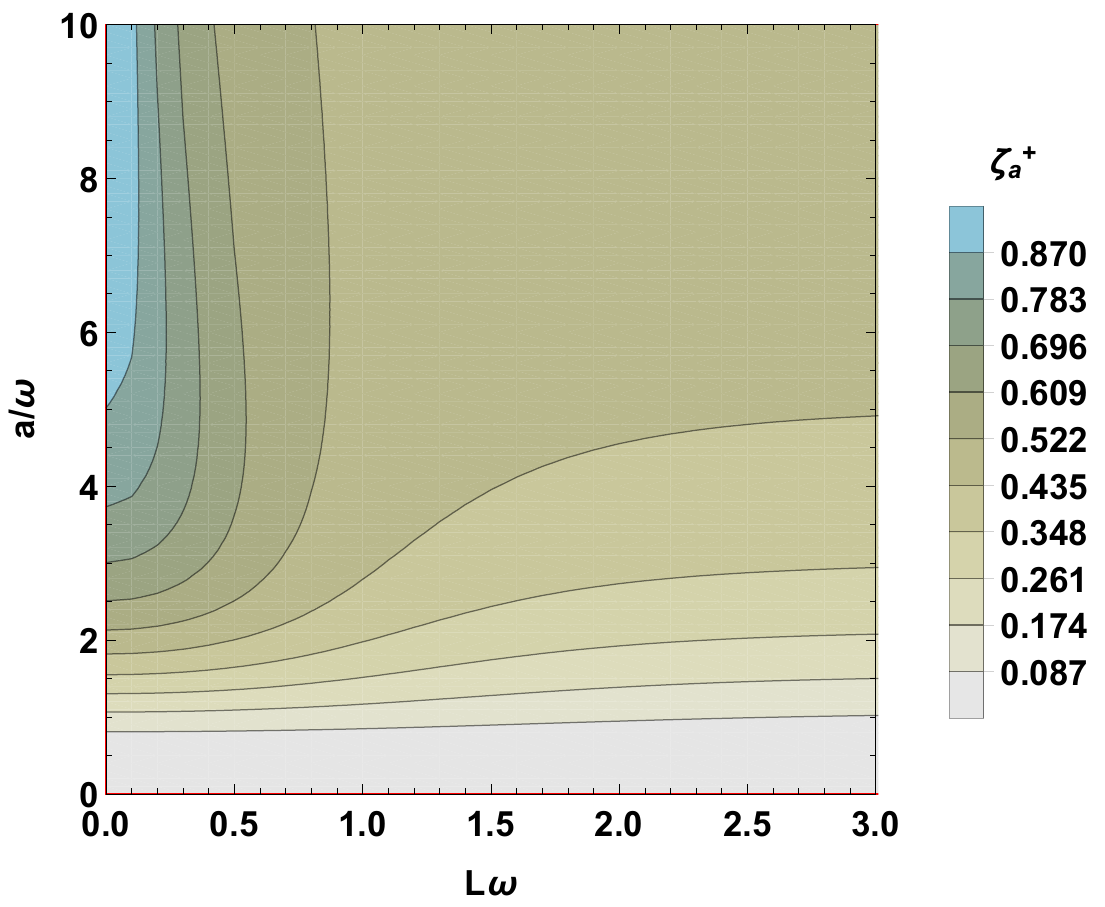}
\caption{\label{zeta(L,a)}
The contour maps of the decoherence factors $\zeta^{-}_{a}$ (left) and $\zeta^{+}_{a}$ (right) in parameter space $(L\omega,a/\omega)$ for two accelerated Unruh-DeWitt detectors coupled with massive scalar fields ($m/\omega=1$).}
\end{figure}

When the acceleration approaches zero, both the decoherence factors $\zeta^{\pm}_{a}$ and $\zeta^{\pm}_{\beta}$ are vanishing. (See Appendix \ref{the properties zeta} for the details of the properties of $\zeta^{\pm}_a$).
In Fig. \ref{zeta(L,a)}, we plot the contour maps of the decoherence factors $\zeta^{-}_{a}$ (left) and $\zeta^{+}_{a}$ (right) in the parameter space $(L\omega,a/\omega)$ for two accelerated Unruh-DeWitt detectors coupled with massive scalar fields.
It can be found that, the factor $\zeta^{-}_{a}$ increases monotonically with acceleration, while  the relation between the factor $\zeta^{+}_{a}$ and acceleration is non-monotonic when the separation of the two detectors is relatively small compared with the transition wavelength.

\paragraph{The mass effects}

For detectors coupled with massless scalar fields, the decoherence factors  in the acceleration case $\zeta^{\pm}_a$ and the thermal bath case $\zeta^{\pm}_\beta$ can be respectively written as
\bea
\zeta^{\pm}_{a}=\frac{\tanh\left(\pi\omega/a\right)}{\pi\omega/a}\frac{1}{2}\left[1\pm\frac{2\sinh^{-1}\left({a L}/{2}\right)}{a L\sqrt{1+a^2L^2/4}}\right],
\;\;\;\zeta^{+}_{\beta}=\frac{\tanh\left(\beta\omega/2\right)}{\beta\omega/2},\;\;\zeta^{-}_{\beta}=0.
\eea
When the fields the detectors coupled to are changed from massless ones to  massive ones, the decoherance factor $\zeta^{\pm}$ disappears for the thermal bath case, but not for the acceleration case. In particular, we obtain that (See Appendix \ref{the properties zeta} for details)
\begin{eqnarray}
&&\lim\limits_{m\rightarrow\infty}\zeta^{+}_a(m/\omega,a/\omega,L\omega)=
\begin{cases}
{\text{sech}(\pi\omega/a)},& L=0,\\
\frac{1}{2}\text{sech}(\pi\omega/a),& L\neq0,
\end{cases}\\
&&\lim\limits_{m\rightarrow\infty}\zeta^{-}_a(m/\omega,a/\omega,L\omega)=
\begin{cases}
0,& L=0,\\
\frac{1}{2}\text{sech}(\pi\omega/a),& L\neq0.
\end{cases}
\end{eqnarray}
That is, even if the mass of the field goes to infinity, the decoherance factors $\zeta^{\pm}_{a}$ are nonzero as long as the separation between the detectors is non-vanishing. To show more details, in Fig. \ref{zeta(L,m)}, we plot the contour maps of the decoherence factors $\zeta^{-}_{a}$ (left) and $\zeta^{+}_{a}$ (right) in the parameter space $(L\omega,m/\omega)$ for two accelerated Unruh-DeWitt detectors coupled with massive scalar fields.
\begin{figure}[!htbp]
\centering
\includegraphics[width=0.495\textwidth]{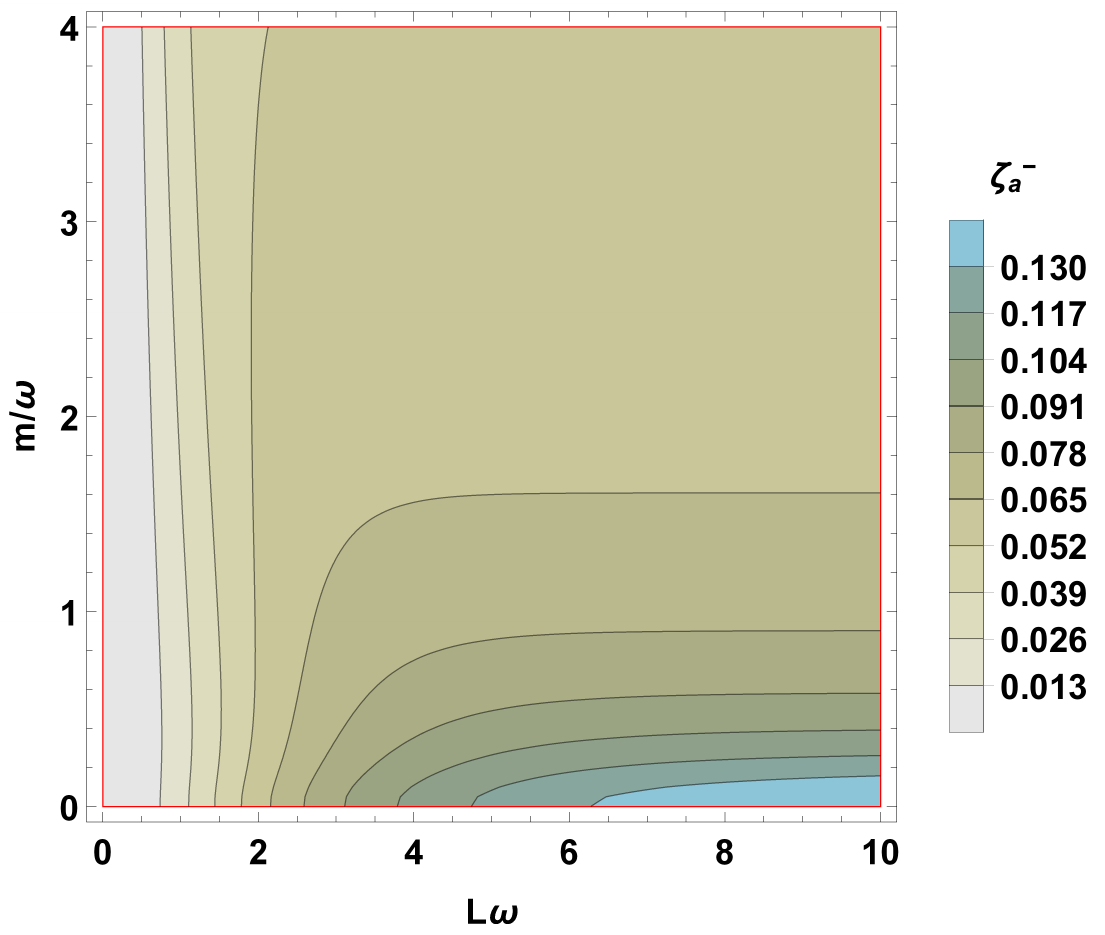}
\includegraphics[width=0.495\textwidth]{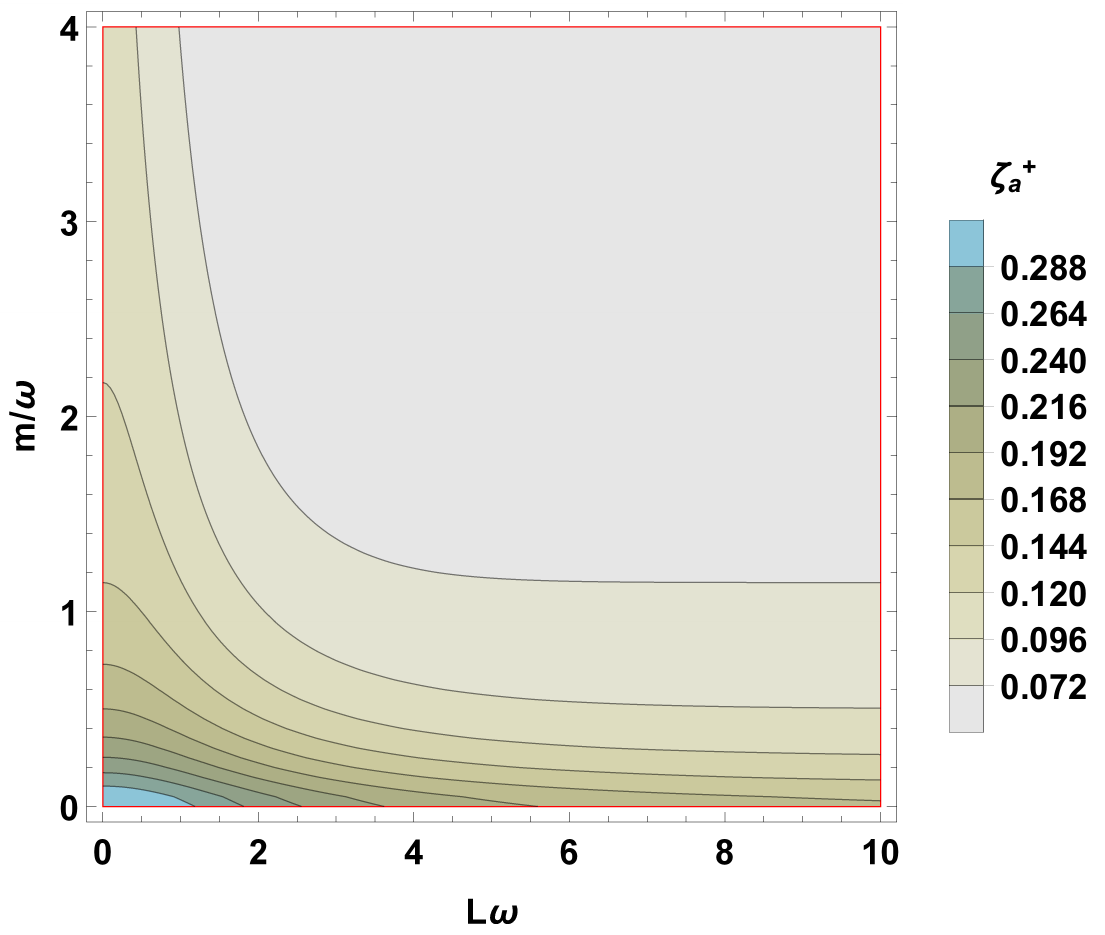}
\caption{\label{zeta(L,m)}
The contour maps of the decoherence factors $\zeta^{-}_{a}$ (left) and $\zeta^{+}_{a}$ (right) in parameter space of $(L\omega,m/\omega)$ for two accelerated ($a/\omega=1$) Unruh-DeWitt detectors coupled with massive scalar fields.}
\end{figure}
As shown in Fig. \ref{zeta(L,m)}, the factor $\zeta^{+}_{a}$ decreases monotonically with the mass of the field, while the relation between the factor $\zeta^{-}_{a}$ and the mass of the field is monotonic only when the separation between the detectors is large enough. This indicates that the mass of the field must weaken the additional entanglement degradation effect brought about by the factor $\zeta^{+}_{a}$;  however, it can both strengthen and weaken the effect brought about by the factor $\zeta^{-}_{a}$.

In conclusion, accelerated detectors not only suffer  the same dissipation effect as that caused by a thermal bath, but also an additional decohernece effect, both of  which contribute to entanglement degradation.

\subsection{The entanglement generation effect}

The entanglement generation effect is crucially dependent on the factor $\lambda$ ($|\lambda|\leq1$) defined in Eq. \eqref{lambdad}.
When $\lambda=0$, it can be obtained from Eq. \eqref{derivatives of K12} that $d C(\tau)/d \tau\leq0$. That is, when $\lambda=0$, it is impossible to create entanglement when the system is initially separable, and the entanglement can only decrease during evolution when the system is initially entangled. In fact, from Eq. \eqref{derivatives of K12}, one can find that a large enough $|\lambda|$ is a necessary condition for entanglement generation.

In the following, we investigate the entanglement generation of a quantum system with the initial state $|E\rangle$. We focus on the relationship between entanglement creation and the factor $\lambda$. In Fig. \ref{Cmax(eta,lambda)}, we show the  maximum of concurrence $C_{max}$ during evolution in the parameter space $(\lambda ,\eta)$ with different decoherence factor $\zeta^{-}$. Here only the $\lambda>0$ part is shown, because it is symmetric about $\lambda=0$. It is shown that, for each dissipative factor $\eta$ and decoherence factor $\zeta^{-}$, there exists a range of $\lambda\in(\lambda_{min},\lambda_{max})$ within which  entanglement can be created. For the acceleration case, this range is related to the field mass $m$, the acceleration $a$ and the separation $L$, while for the thermal case, it is related to the temperature of the bath $T$ only.
One can find from Fig. \ref{Cmax(eta,lambda)} that, the lower  limit $\lambda_{min}$ increases as $\eta$ and $\zeta^{-}$ increase.
\begin{figure}[!htbp]
\centering
\includegraphics[width=0.328\textwidth]{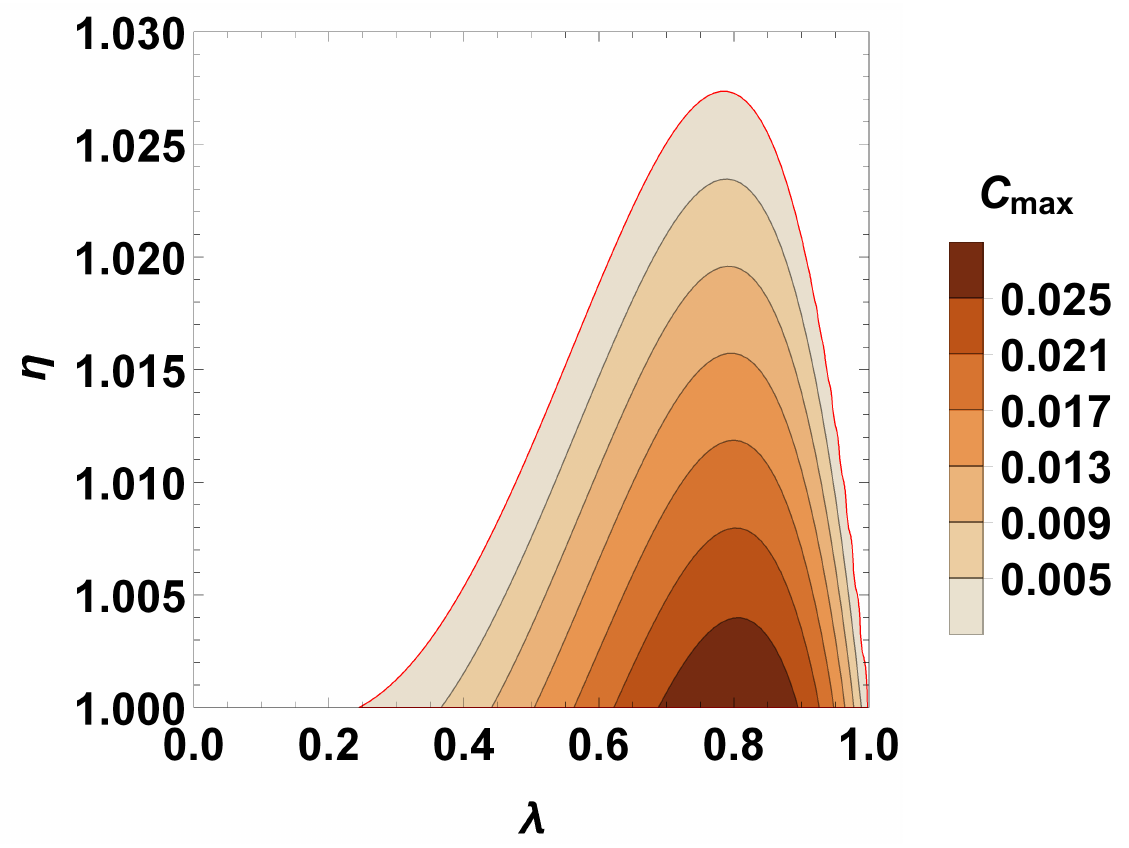}
\includegraphics[width=0.328\textwidth]{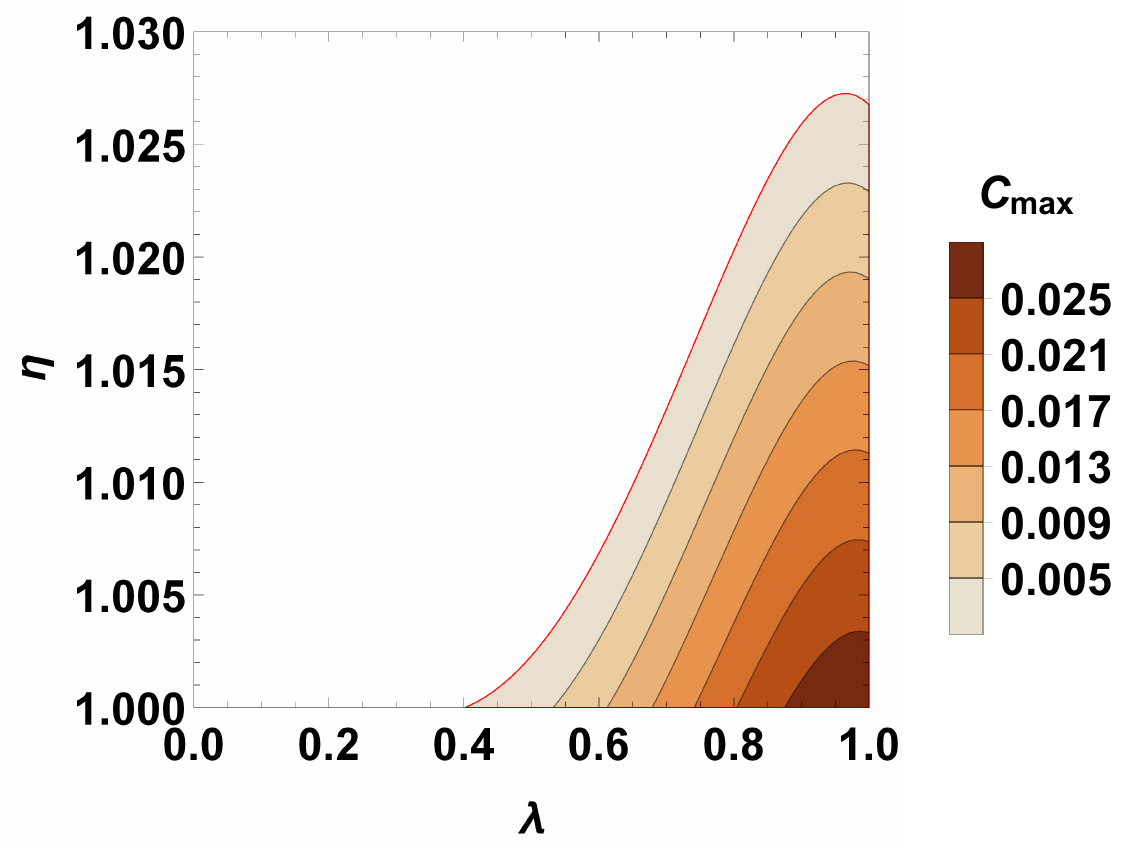}
\includegraphics[width=0.328\textwidth]{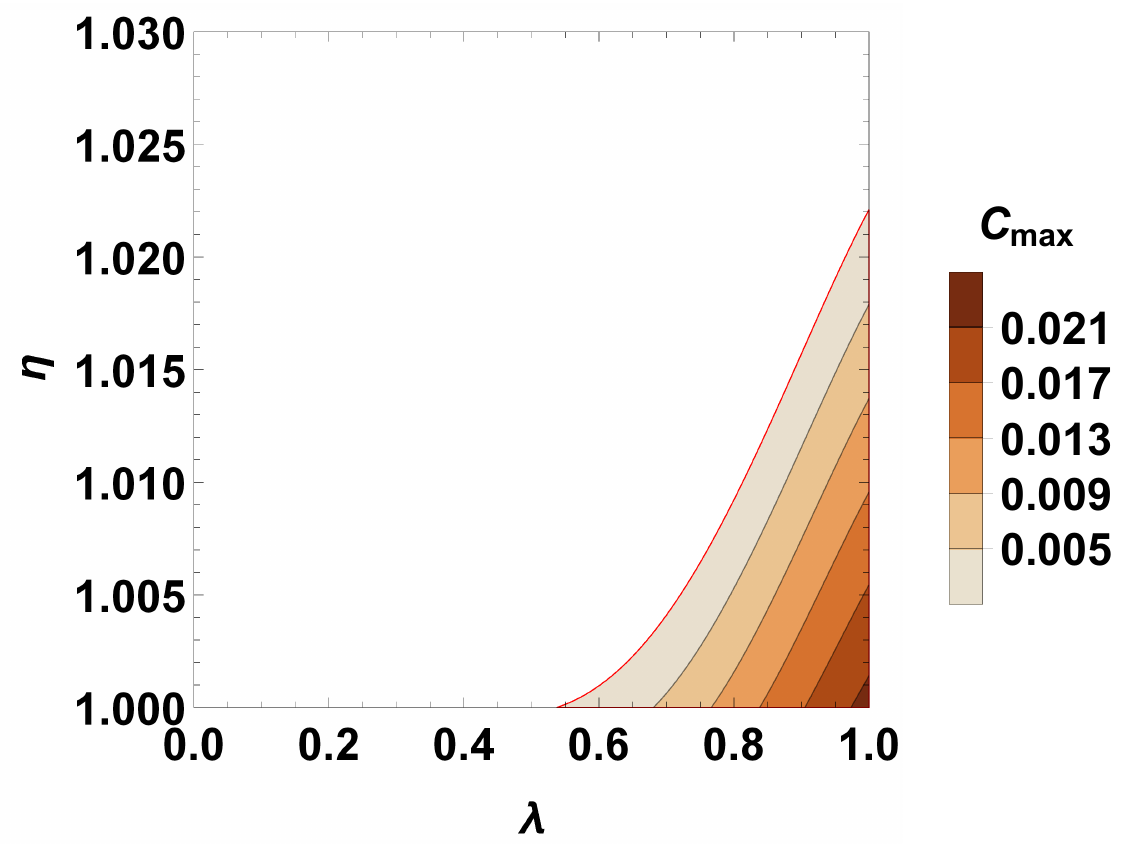}
\caption{\label{Cmax(eta,lambda)}
The contour maps of the maximum of concurrence during evolution $C_{max}$ in parameter space $(\lambda,\eta)$ with $\zeta^{-}=0$ (left), 0.1 (middle), and 0.2 (right) for quantum systems initially prepared in $|E\rangle$. Note that contour lines with $C_{max}<0.001$ are not drawn here.}
\end{figure}

For a quantum system composed of two uniformly accelerated Unruh-DeWitt detectors coupled with  massive scalar field in the Minkowski vacuum, the coherence factor $\lambda_a$  can be written as $\lambda_a(m/\omega,a/\omega,L\omega)$, whose explicit expression has been given in Eq. (\ref{lambda a}), and the expressions in several limiting cases have been shown in Appendix \ref{the properties lambda}. Now, we focus on how the mass of the field and the acceleration affect the parameter regions within which the system can get entangled.

\subsubsection{The mass effects}

In this part, we  study the  parameter regions within which entanglement creation can occur for two uniformly accelerated Unruh-DeWitt detectors prepared in $|E\rangle$ coupled with massive scalar fields, and compare the result with  that  in  the  massless case. We  examine two situations, namely $m\geq\omega$ (Fig. \ref{lambda(L,a)m08mn}) and $m<\omega$ (Fig. \ref{lambda(L,a)m12mn}).

\begin{figure}[!htbp]
\centering
\begin{tikzpicture}
\scope[nodes={inner sep=0,outer sep=0}]
\node[anchor=north east] (a)
 {\includegraphics[width=6cm]{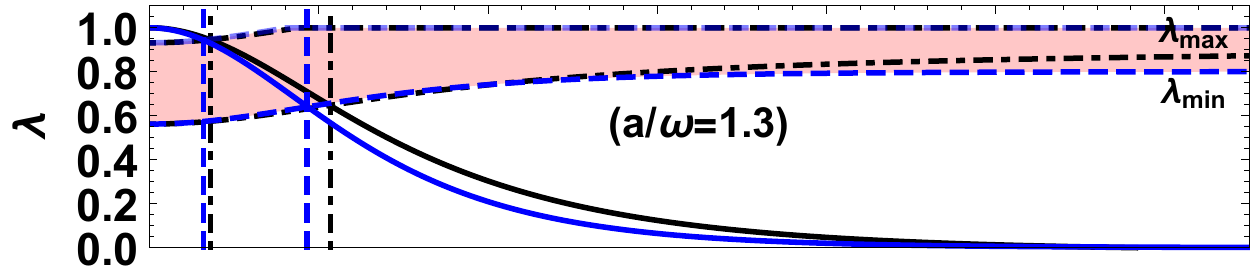}};
\node[below=0mm of a.south west,anchor=north west] (b)
{\includegraphics[width=6cm]{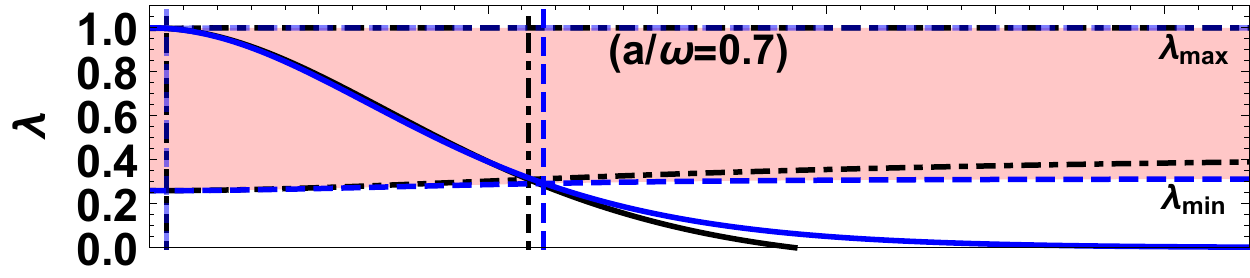}};
\node[below=0mm of b.south west,anchor=north west] (c)
{\includegraphics[width=6cm]{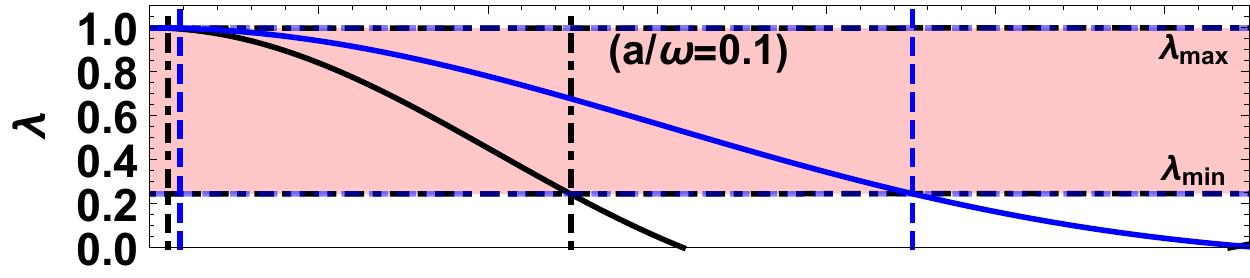}};
\node[below=0mm of c.south west,anchor=north west] (d)
{\includegraphics[width=6cm]{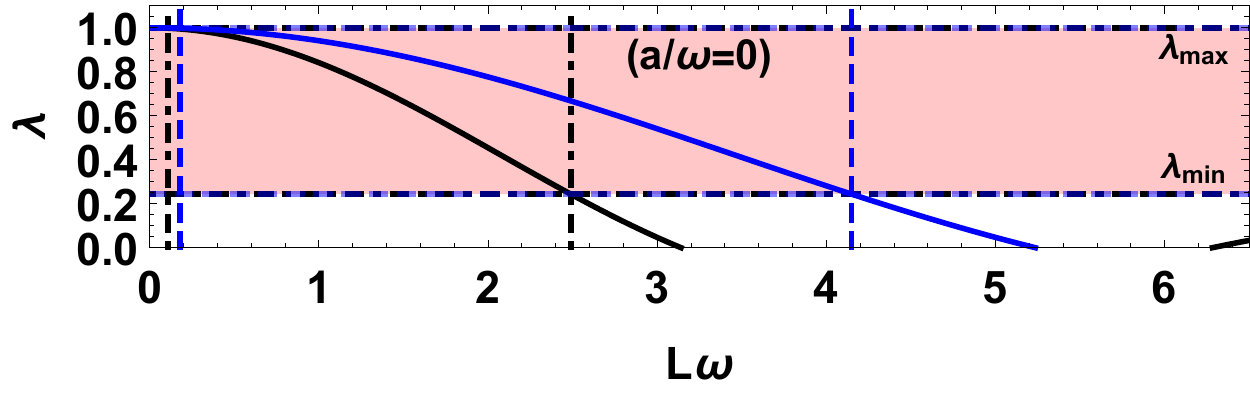}};
\node[right=6mm of a.north east,anchor=north west] (e)
{\includegraphics[height=6cm]{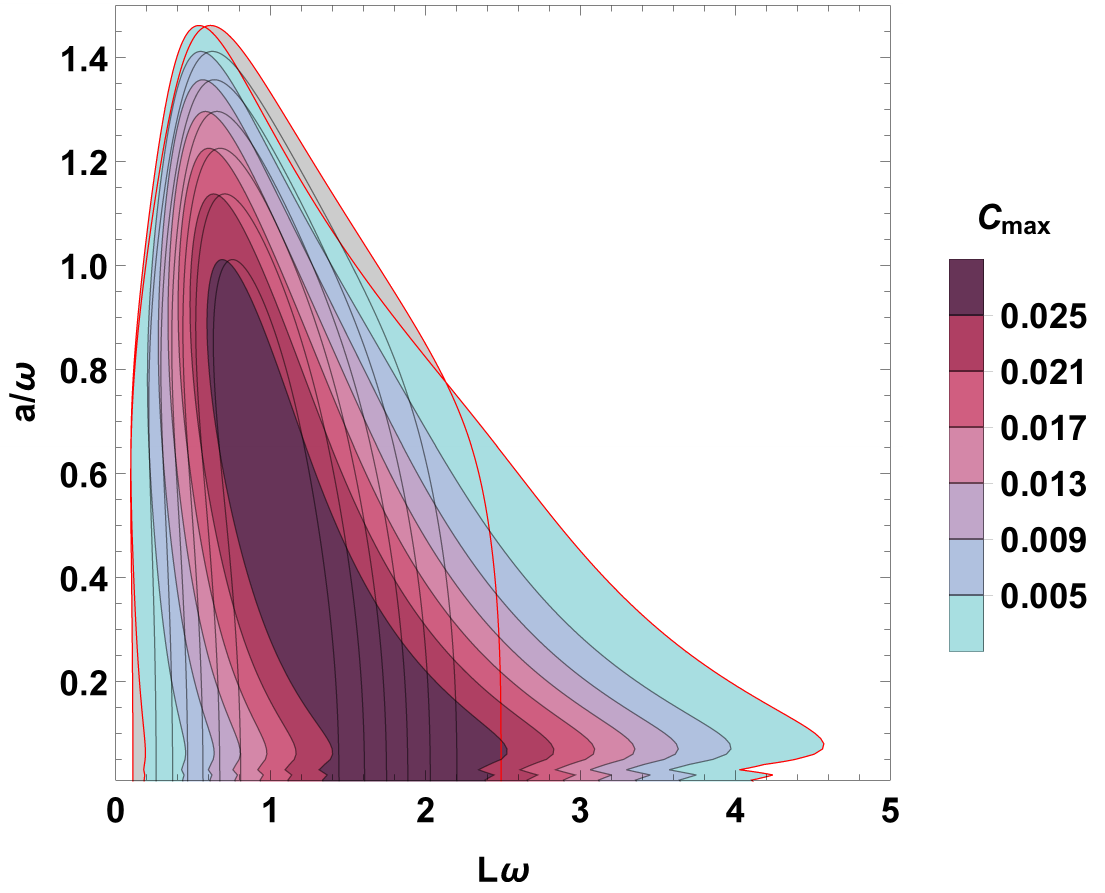}};
\endscope
\end{tikzpicture}
\caption{\label{lambda(L,a)m08mn}
(Left)  The factor $\lambda$ as a function of separation $L\omega$, with different accelerations $a/\omega$. The horizontal blue-dashed and black-dot-dashed lines mark the upper and lower limits $\lambda_{max}$ and $\lambda_{min}$ within which entanglement can be generated ($C_{max}\geq0.001$).
(Right) Contour map of the maximal  concurrence  during evolution for quantum systems prepared in $|E\rangle$   in the  parameter space $(L\omega, a/\omega)$. The figures show the differences between the results in the massive case ($m/\omega=0.8$, the blue lines on the left,  and the colored contour on the right) and the  massless case (the black lines on the left, and the un-colored contour on the right).  }
\end{figure}

Firstly, in Figs. \ref{lambda(L,a)m08mn} and \ref{lambda(L,a)m12mn}, it is shown that, compared with the massless case, the region of separation $L\omega$ within which  entanglement can be created is expanded  when the acceleration $a/\omega$ is smaller than a critical value $a_{crit}/\omega$ determined by the field mass $m/\omega$, and is compressed when $a>a_{crit}$. This is distinct from the fact that the region of separation for entanglement generation  is always expanded for a static quantum system coupled with massive fields (in  vacuum or in a thermal bath), compared with the massless case \cite{Zhou2020}.
\begin{figure}[!htbp]
\centering
\begin{tikzpicture}
\scope[nodes={inner sep=0,outer sep=0}]
\node[anchor=north east] (a)
 {\includegraphics[width=6cm]{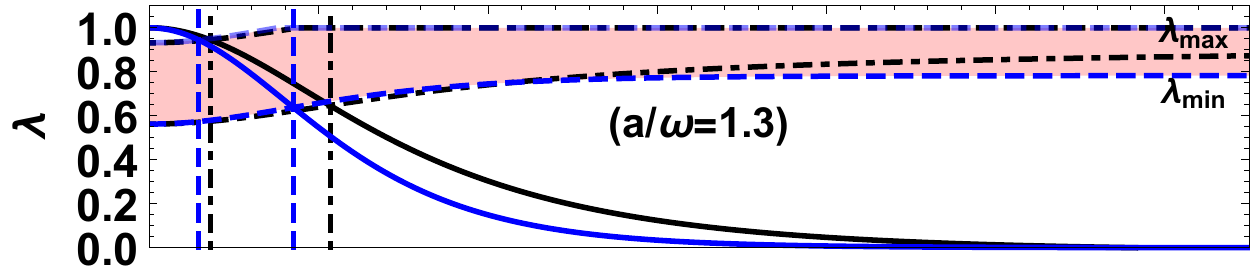}};
\node[below=0mm of a.south west,anchor=north west] (b)
{\includegraphics[width=6cm]{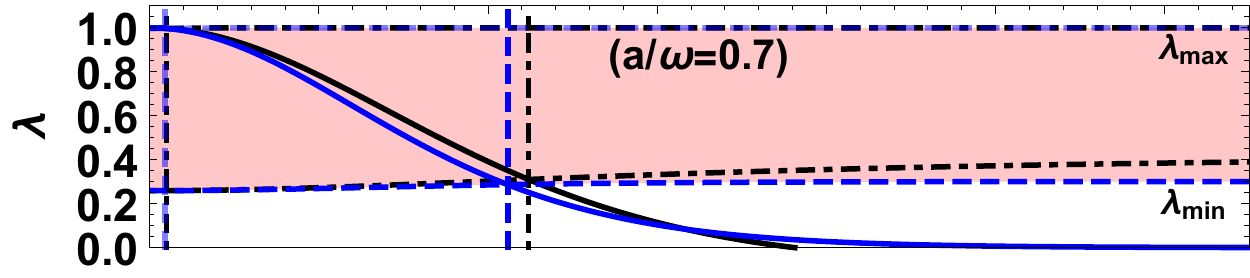}};
\node[below=0mm of b.south west,anchor=north west] (c)
{\includegraphics[width=6cm]{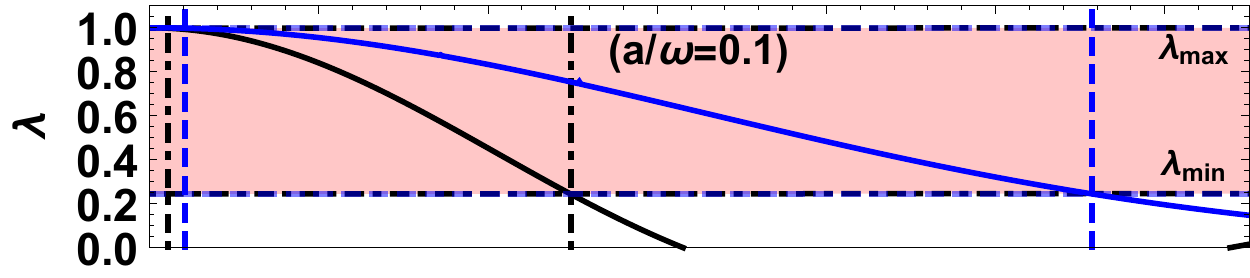}};
\node[below=0mm of c.south west,anchor=north west] (d)
{\includegraphics[width=6cm]{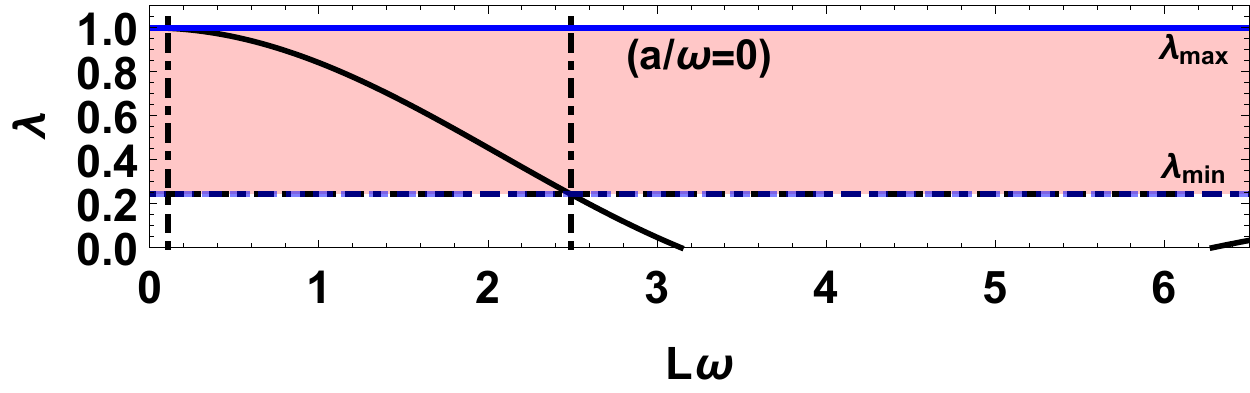}};
\node[right=6mm of a.north east,anchor=north west] (e)
{\includegraphics[height=6cm]{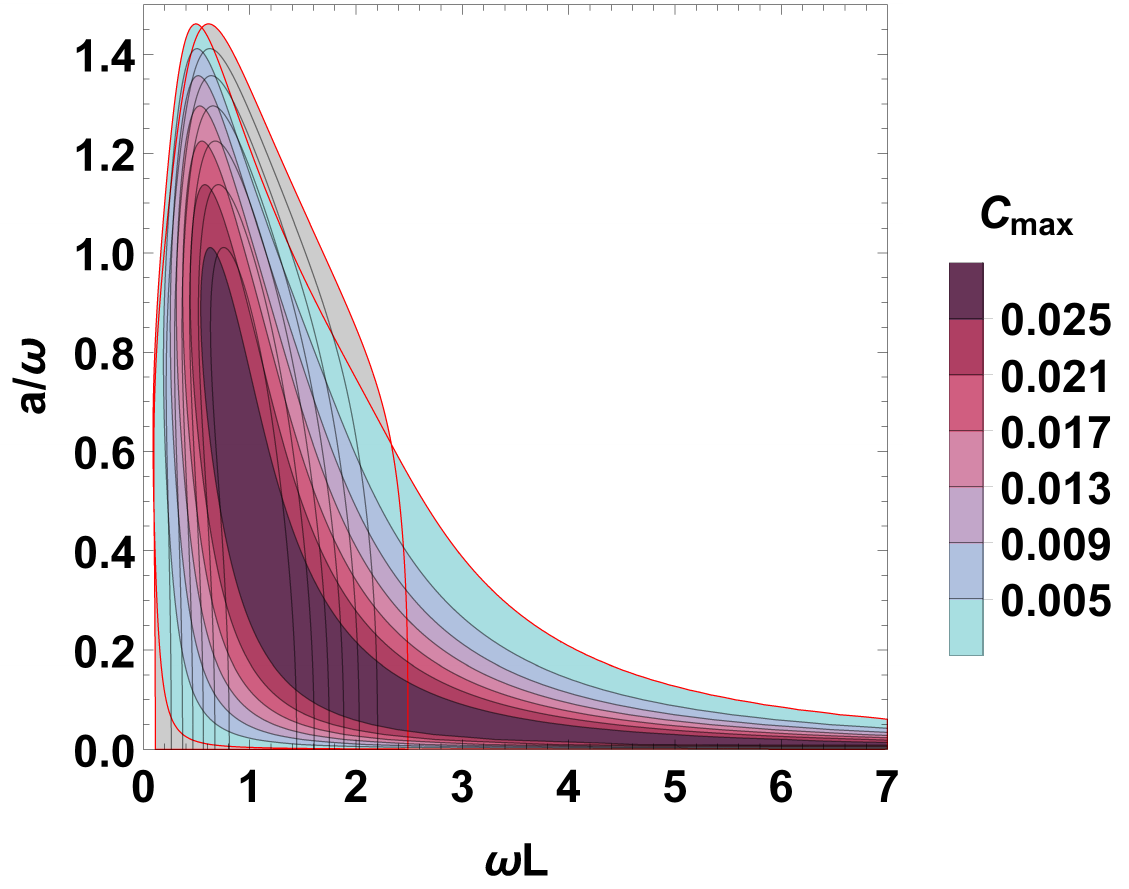}};
\endscope
\end{tikzpicture}
\caption{\label{lambda(L,a)m12mn}
(Left)  The factor $\lambda$ as a function of separation $L\omega$, with different accelerations $a/\omega$. The horizontal blue-dashed and black-dot-dashed lines mark the upper and lower limits $\lambda_{max}$ and $\lambda_{min}$ within which entanglement can be generated ($C_{max}\geq0.001$). (Right) Contour map of the maximal  concurrence  during evolution for quantum systems prepared in $|E\rangle$   in  the parameter space $(L\omega, a/\omega)$. The figures show the differences between the results in the massive case ($m/\omega=1.2$, the blue lines on the left,  and the colored contour on the right) and the  massless case (the black lines on the left, and the un-colored contour on the right).}
\end{figure}

Secondly, by comparing Fig. \ref{lambda(L,a)m08mn} and Fig. \ref{lambda(L,a)m12mn}, we find that the behaviors of the factor $\lambda_a$ at $a\rightarrow0$ is completely different for $m<\omega$ and $m>\omega$, resulting in a significant difference in the possible regions of separation $L\omega$ for entanglement creation. In the acceleration case, when the mass of the field exceeds the energy level spacing of the Unruh-DeWitt detectors, long-distance entanglement creation can  be achieved  when the acceleration is small, in contrast to the static case in which the mass of the field has to be smaller than but close to the energy level spacing in order to  achieve long-distance entanglement  \cite{Zhou2020}.

\subsubsection{The acceleration effects}
Now  we focus on the effects of acceleration on entanglement evolution, and  compare the results with those of static ones in a thermal bath at the Unruh temperature. First, according to the analytical expressions Eqs. \eqref{lambda a} to \eqref{lambda b}, in the acceleration case, the  factor $\lambda_a$ is a function of the acceleration $a$, while in the thermal case, $\lambda_{\beta}$ is independent of  temperature, which results in  essential differences between the two cases.

\begin{figure}[!htbp]
\centering
\begin{tikzpicture}
\scope[nodes={inner sep=0,outer sep=0}]
\node[anchor=north east] (a)
 {\includegraphics[width=6cm]{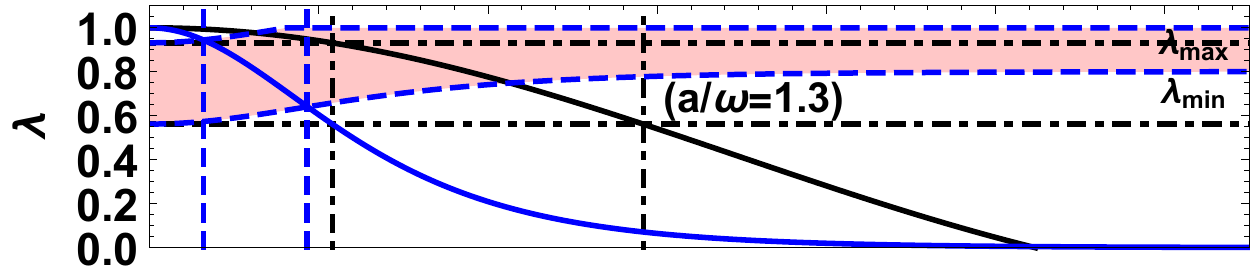}};
\node[below=0mm of a.south west,anchor=north west] (b)
{\includegraphics[width=6cm]{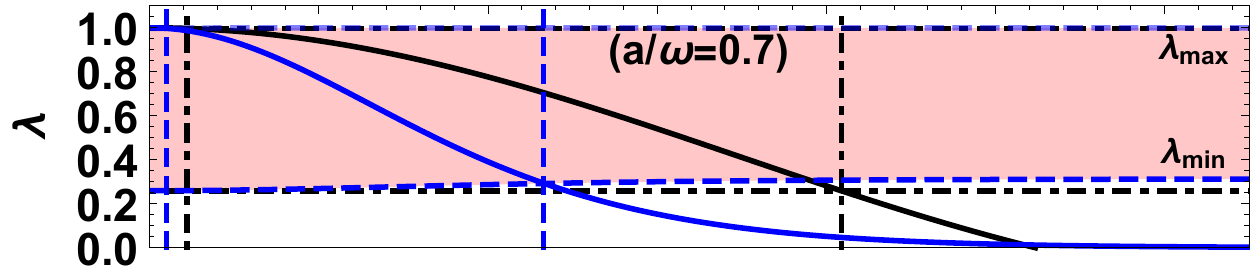}};
\node[below=0mm of b.south west,anchor=north west] (c)
{\includegraphics[width=6cm]{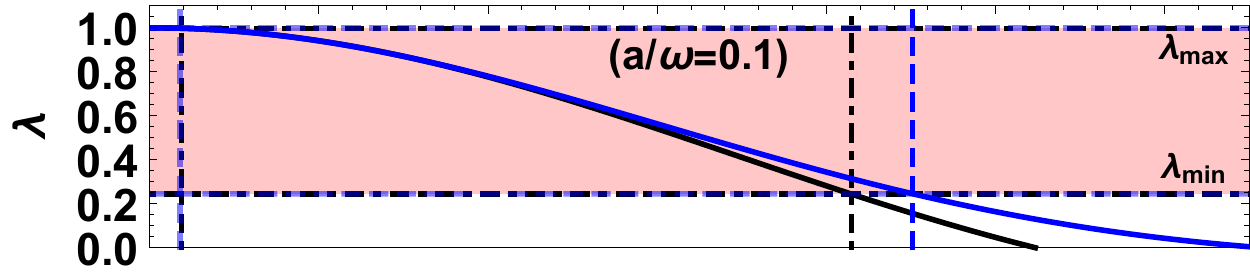}};
\node[below=0mm of c.south west,anchor=north west] (d)
{\includegraphics[width=6cm]{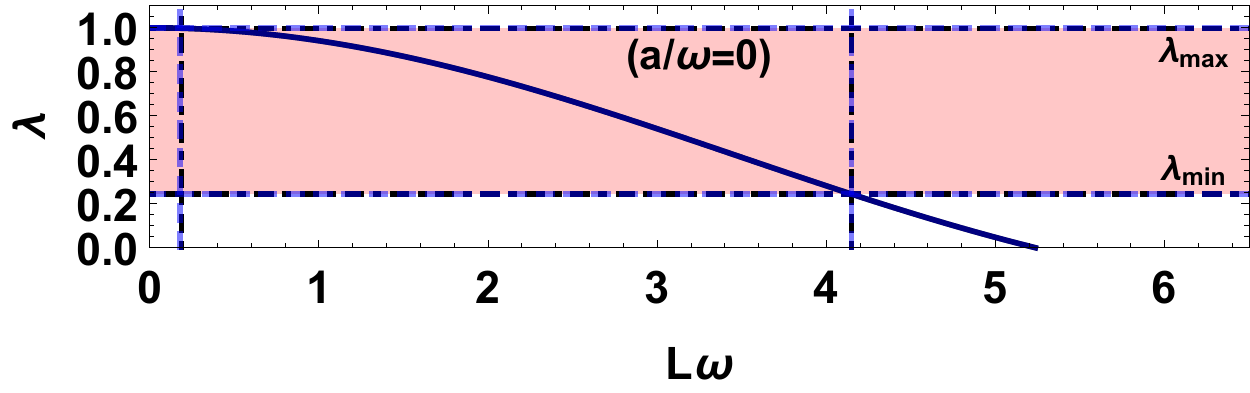}};
\node[right=6mm of a.north east,anchor=north west] (e)
{\includegraphics[height=6cm]{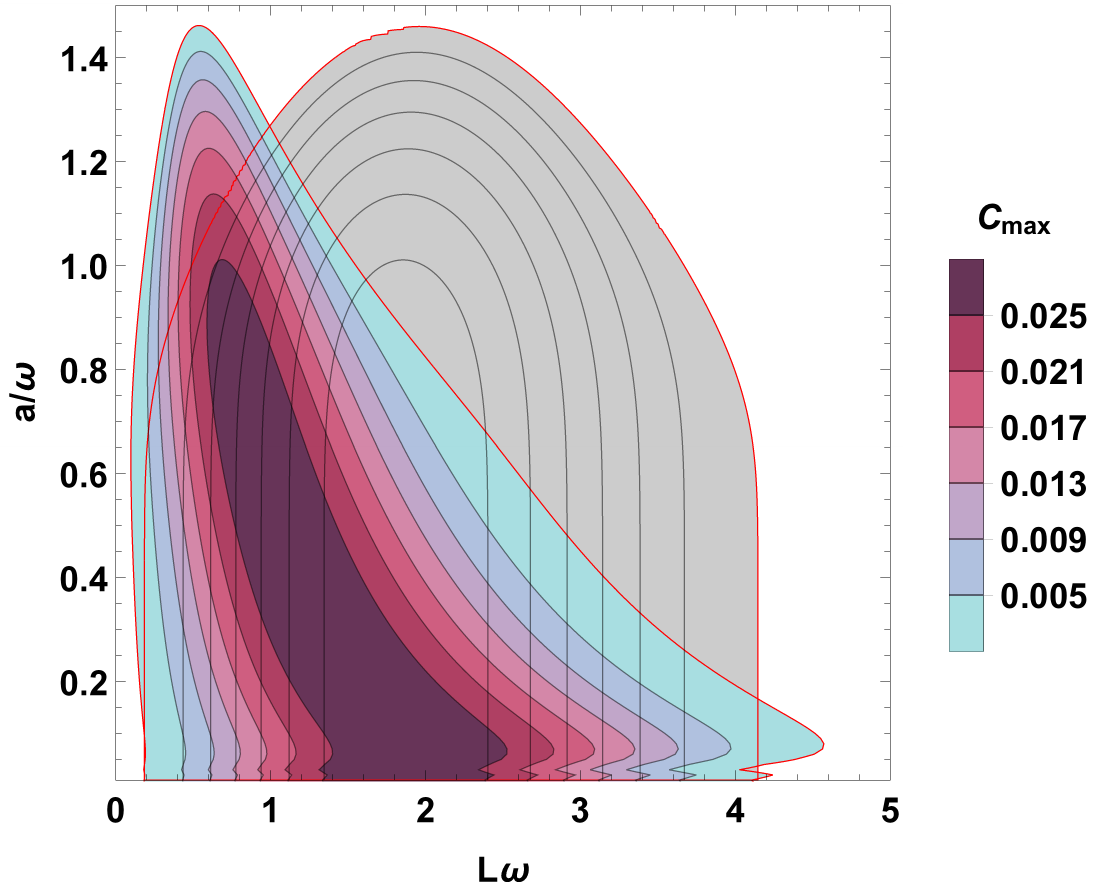}};
\endscope
\end{tikzpicture}
\caption{\label{lambda(L,a)}
(Left) The factor $\lambda$ as a function of  separation $L\omega$, with  different accelerations $a/\omega$.  The horizontal blue-dashed and black-dot-dashed lines mark the upper and lower limits $\lambda_{max}$ and $\lambda_{min}$ within which entanglement can be generated ($C_{max}\geq0.001$). (Right) Contour map of the maximal concurrence during evolution for quantum systems prepared in $|E\rangle$ with $m/\omega=0.8$ in the parameter space $(L\omega, a/\omega)$. The figures show the differences between the results in the  acceleration case (the blue lines on the left, and the colored contour on the right) and the thermal case (the black lines on the left, and the un-colored contour on the right).}
\end{figure}

In Fig. \ref{lambda(L,a)}, it is shown that when $a/\omega=0$, the acceleration case is the same as the static case as expected. When the acceleration increases,  in a certain region where the acceleration is relatively small, the possible region of separation $L\omega$ for entanglement generation  varies  oscillatorily   from being compressed to being enlarged compared with that in the thermal case, due to the oscillation of $\lambda_a$ with  acceleration $a$.
The amplitude of this oscillation increases with acceleration, until the maximal amplitude is reached (at about $a/\omega\approx 0.1$ in Fig. \ref{lambda(L,a)}),  which causes a maximal expansion of the possible region of separation $L\omega$ for entanglement generation. Then,  $\lambda_a$ decays monotonically  with  acceleration, which causes the compression of the possible region of separation $L\omega$ for entanglement generation compared with that of the thermal case, as shown in Fig. \ref{lambda(L,a)}.
However, in the thermal case, since $\lambda_{\beta}$ is independent of  temperature,
the possible region of separation $L\omega$ for entanglement generation decreases monotonically as temperature increases.

\begin{figure}[!htbp]
\centering
\begin{tikzpicture}
\scope[nodes={inner sep=0,outer sep=0}]
\node[anchor=north east] (a)
 {\includegraphics[width=6cm]{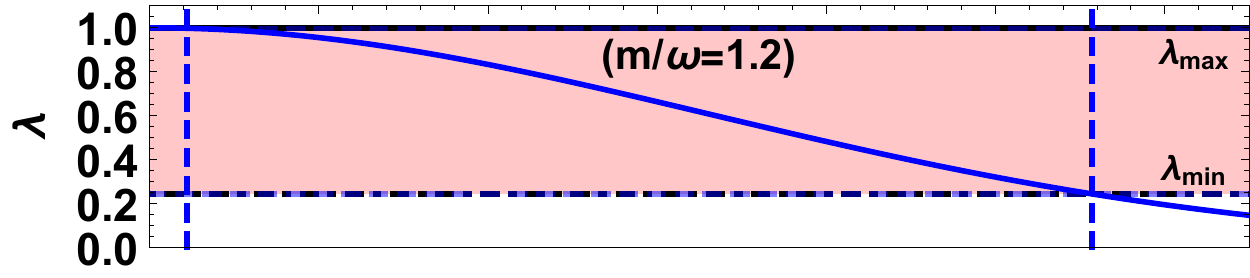}};
\node[below=0mm of a.south west,anchor=north west] (b)
{\includegraphics[width=6cm]{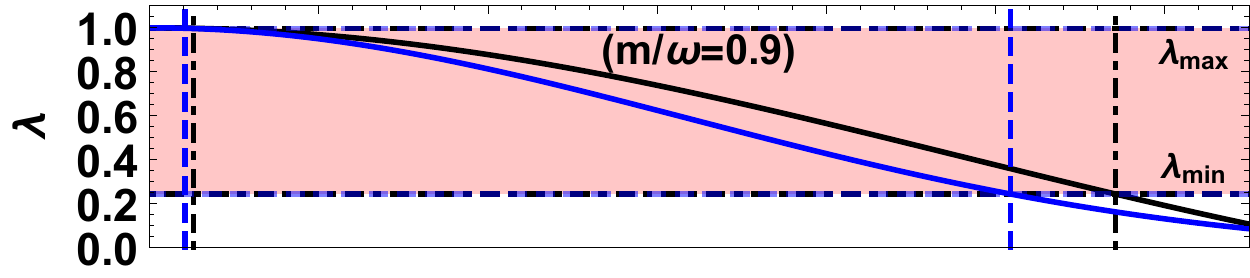}};
\node[below=0mm of b.south west,anchor=north west] (c)
{\includegraphics[width=6cm]{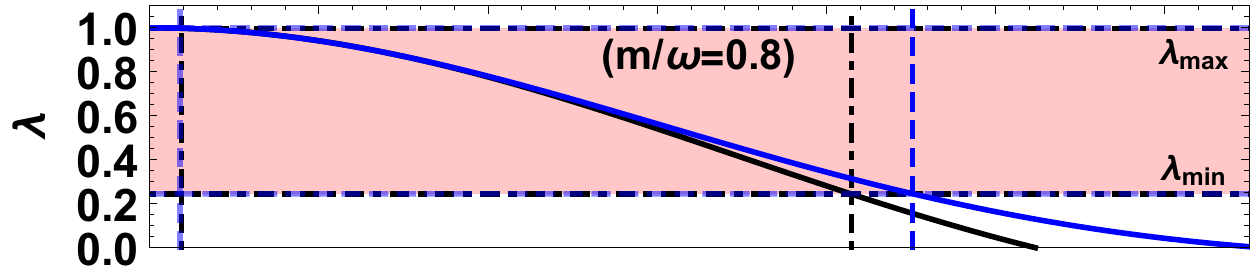}};
\node[below=0mm of c.south west,anchor=north west] (d)
{\includegraphics[width=6cm]{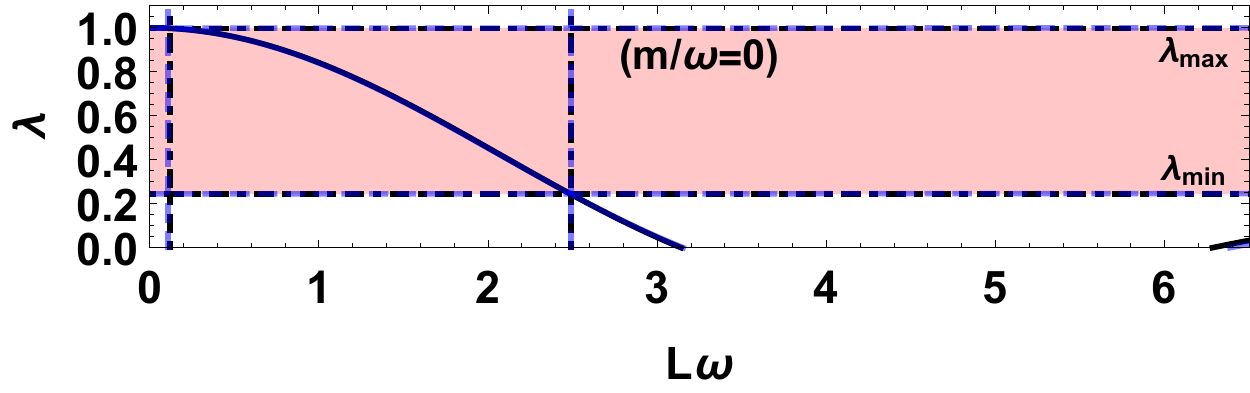}};
\node[right=6mm of a.north east,anchor=north west] (e)
{\includegraphics[height=6cm]{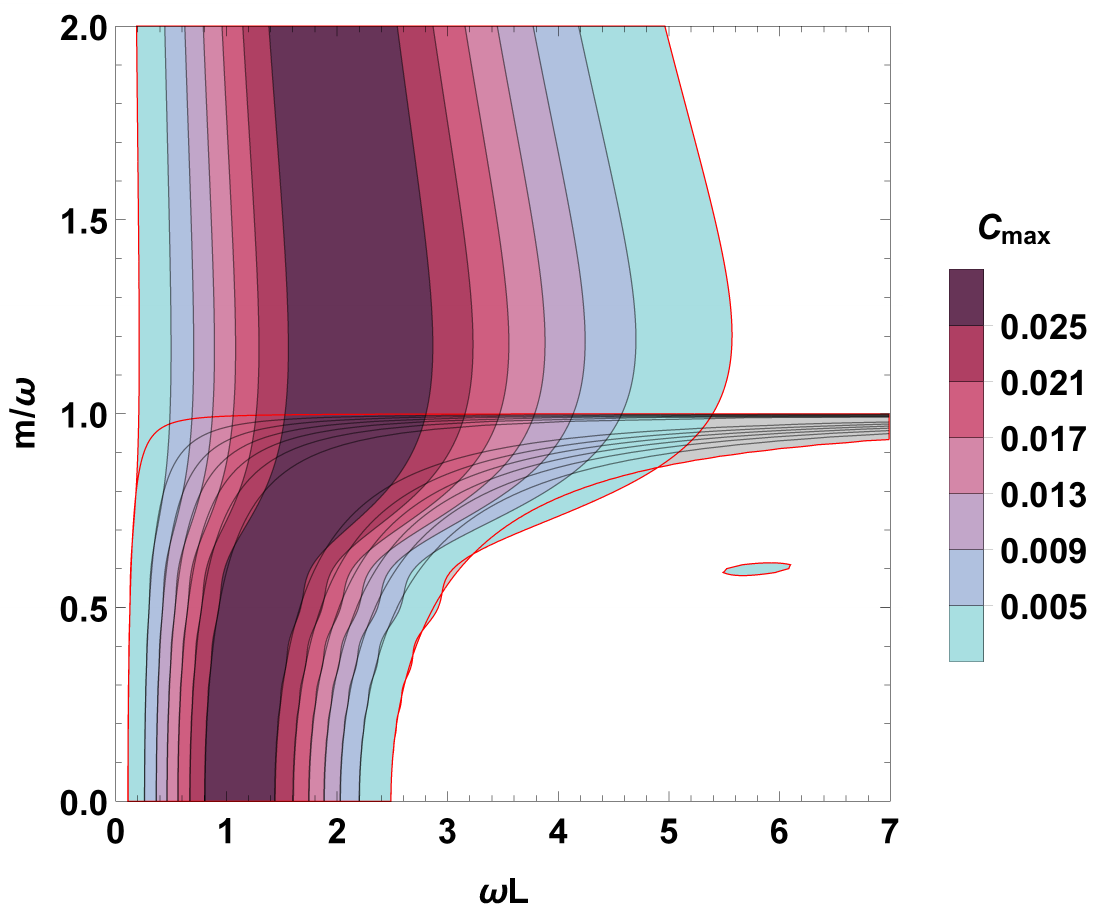}};
\endscope
\end{tikzpicture}
\caption{\label{lambda(L,m)}
(Left) The factor $\lambda$ as a function of  separation $L\omega$, with  different  mass  $m/\omega$. The horizontal blue-dashed and black-dot-dashed lines mark the upper and lower limits $\lambda_{max}$ and $\lambda_{min}$ within which entanglement can be generated ($C_{max}\geq0.001$).  (Right) Contour map of the maximal concurrence during evolution for quantum systems prepared in $|E\rangle$ with  $a/\omega=0.1$  in the parameter space $(L\omega, m/\omega)$. The figures show the differences between the results in the  acceleration case  (the blue lines on the left, and the colored contour on the right) and the  thermal case (the black lines on the left, and the un-colored contour on the right).  }
\end{figure}

In Fig. \ref{lambda(L,m)}, we study the effects of mass on entanglement dynamics for two uniformly accelerated Unruh-DeWitt detectors, and compare the results with those of  in the thermal case. From  Fig. \ref{lambda(L,m)}, we  draw the following conclusions\footnote{In Fig. \ref{lambda(L,m)}, the acceleration is set to $a/\omega=0.1$. However, the conclusions still hold if a larger acceleration is chosen.}:

\emph{1)}
When $m/\omega<1$,
in the  thermal case, the damping of $\lambda_{\beta}$ with $L\omega$ gradually slows down as $m/\omega$ increases. When $m/\omega \to 1$, $\lambda_{\beta}$ becomes a constant ($\lambda_{\beta}=1$). As a result, the region of $L\omega$ within which entanglement can be generated is significantly  enlarged when $m/\omega$ is close to 1. However, in the acceleration case, $\lambda_a$ does not approach a constant no matter how large the mass is. Therefore, the regions of $L\omega$ within which the two detectors can be entangled are only slightly enlarged.

\emph{2)} When $m/\omega\geq1$, there are significant  differences between the acceleration case and the  thermal case. For the thermal case, the factor $\Omega_\beta$  equals to 0, so the detectors are locked up in the  initial state as if it were a closed system, and  entanglement generation cannot occur.
However, in the  acceleration case, no matter how large $m/\omega$ is (as long as it is not infinite, see Eq. \eqref{limitm2}), $\lambda_a$ is a function of  $L\omega$ ranging from 0 to 1, and  the factor $\Omega_a\neq0$.  Therefore, entanglement generation is possible for certain $L\omega$  (but it may take a long time since the factor $\Omega_a$  decays exponentially as $m/\omega$ increases).  A similar conclusion has been drawn in Ref. \cite{Kaplanek2020}, in which it has been found that a single detector relaxes to its thermal equilibrium state extremely inefficiently in the large mass limit.

\subsection{The Unruh and anti-Unruh effects}

The Unruh effect can be sensed by a quantum system coupled with the vacuum fields, i.e. an Unruh-DeWitt detector \cite{W. G. Unruh,dewitt}. For static detectors in a thermal bath, it is well-known that the higher the temperature, the more often the detector clicks. Since the Unruh temperature is proportional to the proper acceleration \cite{W. G. Unruh}, it is expected the larger the acceleration, the more often the  Unruh-DeWitt detector clicks. However, in certain cases, the  transition rate of a uniformly accelerated detector may decrease with acceleration in some parameter regimes,  e.g. in the presence of a boundary \cite{Lu2005}, when the field the detector coupled with is massive \cite{Y. B. Zhou}, and when the duration of the detector-field coupling is finite  \cite{W. Brenna} (but long enough to satisfy the KMS condition \cite{Kubo1957,Martin1959,Haag1967}). This phenomenon is named as the anti-Unruh effect in Ref. \cite{W. Brenna}, and is further divided into two categories \cite{Anti-Unruh2016}, i.e. the strong anti-Unruh effect (the effective excitation-to-deexcitation ratio (EDR) temperature of a detector decreases as the KMS temperature increases), and the weak anti-Unruh effect (a detector  clicks less often as the KMS temperature increases), which is a necessary condition for the  strong one. Recently, it is found that the anti-Unruh effect introduced in Refs. \cite{W. Brenna,Anti-Unruh2016}  may possibly be viewed as an amplification mechanism for quantum entanglement within a certain parameter regime \cite{Anti-Unruh2018}. Moreover, an anti-Unruh phenomenon in the transition probabilities of two entangled uniformly accelerated atoms in a thermal bath has been shown in  Ref. \cite{S. Barman2021}.

In Ref. \cite{Hu}, we have shown that, for a pair of two-level detectors coupled with massless scalar fields,  the maximal concurrence during evolution may increase with acceleration  for specific inter-detector separations, in contrast to the fact that it always decreases monotonically  with temperature in the thermal case. This may also be called as an anti-Unruh phenomenon in terms of the entanglement generated.
In  the massive case, as shown in the contour maps (Figs. \ref{lambda(L,a)m08mn} and  \ref{lambda(L,a)m12mn}), for any inter-detector separation, the maximal concurrence  quantifying the entanglement generated during evolution may increase with acceleration when the acceleration is relatively small with respect to the energy level spacing of the detectors. Therefore, in contrast to the massless case in which the  anti-Unruh phenomenon appears only for specific separation, it is a  general phenomenon in the massive case.

\begin{figure}[!htbp]
\centering
\includegraphics[width=0.49\textwidth]{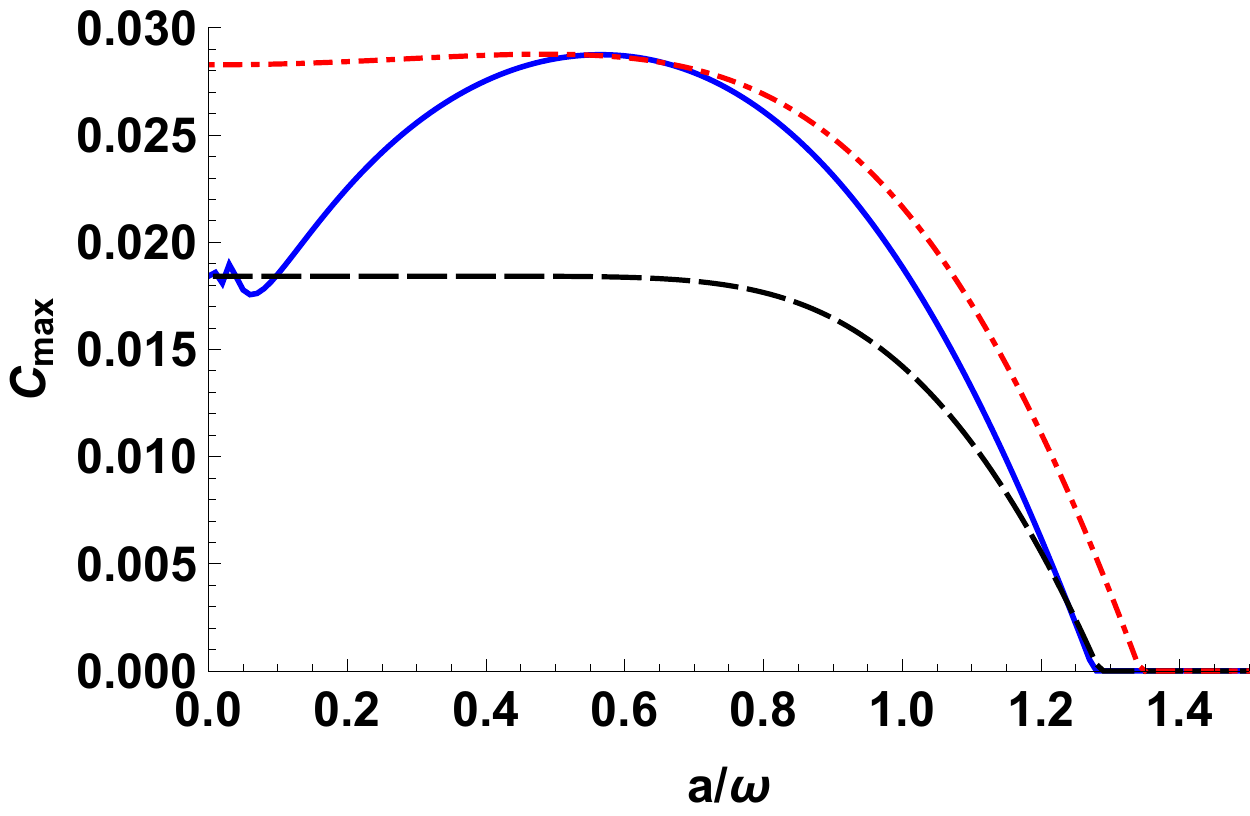}
\includegraphics[width=0.5\textwidth]{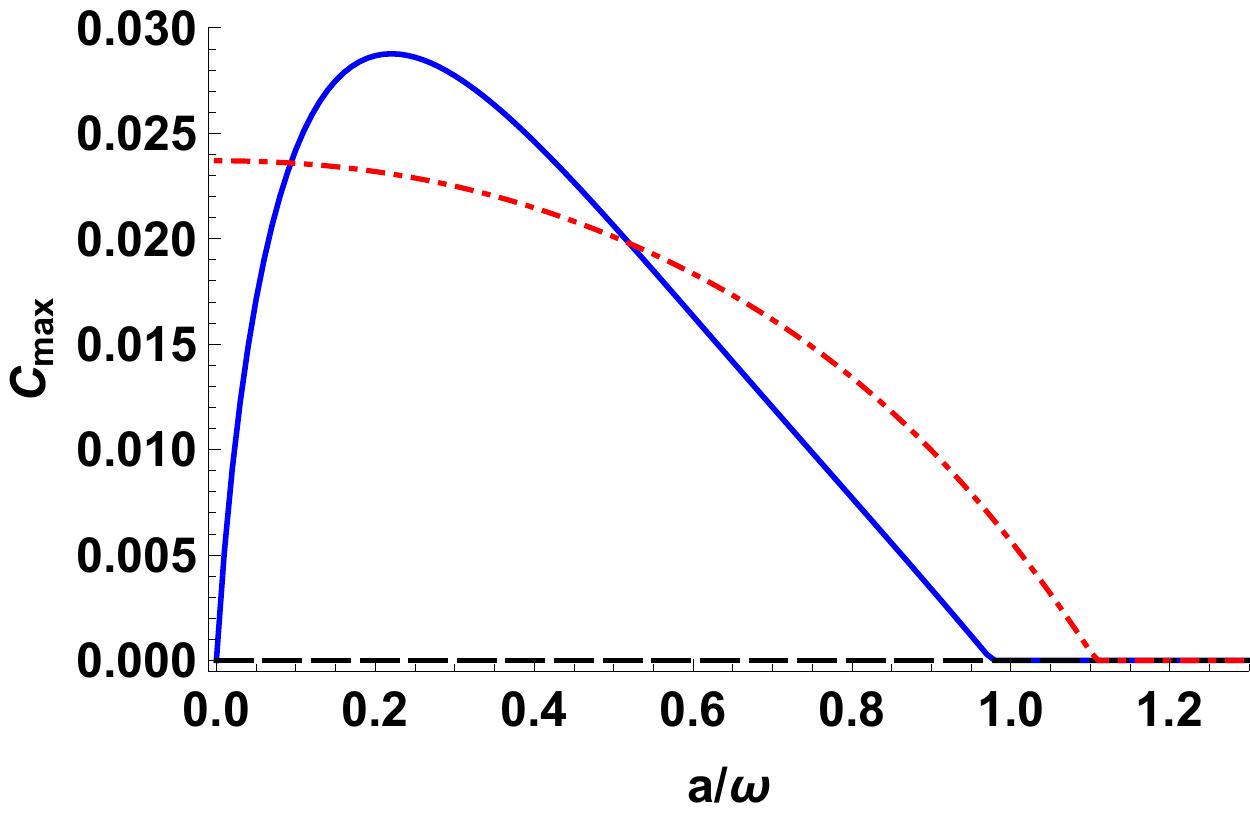}
\caption{\label{Anti-Unruh}
 Comparison between the maximum of concurrence during evolution for uniformly accelerated quantum systems coupled with massive (blue solid lines) and massless (red dot-dashed lines),  and static ones coupled with massive scalar fields in a thermal bath at the Unruh temperature (black dashed lines) initially prepared in $|E\rangle$. Here, $m/\omega=0.8$, $L\omega=1.0$ (left), and $m/\omega=1.2$, $L\omega=1.5$ (right).}
\end{figure}

As an example, in Fig. \ref{Anti-Unruh}, we plot the  maximal concurrence as a function of acceleration $a/\omega$ with fixed mass $m/\omega$ and separation $L\omega$. As shown in Fig. \ref{Anti-Unruh} (left), when $m/\omega<1$, for small accelerations ($a/\omega<0.57$), the maximal concurrence oscillates with acceleration since the entanglement creation effect from $\lambda$ which oscillates with acceleration is  stronger than the entanglement degradation effect from $\eta$. However, when the acceleration is large enough ($a/\omega>0.57$), the maximal concurrence decreases monotonically with acceleration.
When $m/\omega>1$, since $\lambda$ decays monotonically with acceleration from $1$ (when $a/\omega=0$) to $0$ (when $a/\omega\rightarrow\infty$),  the maximal concurrence varies from $0$ (when $a/\omega=0$) to a certain non-zero maximum value (when $a/\omega=0.22$), and then to $0$ (when $a/\omega>0.98$), as shown in Fig. \ref{Anti-Unruh} (right). In either case, the anti-Unruh phenomenon occurs when the acceleration is relatively small with respect to the energy level spacing of the detectors ($a/\omega<0.57$ in the first case, and $a/\omega<0.22$ in the second case).

Based on the previous discussions, the Unruh and anti-Unruh effect in terms of entanglement dynamics can be understood as follows.

Firstly, from the perspective of dissipation (related to $\eta$), an accelerated quantum system in the  Minkowski vacuum suffers  the same dissipative effect as that of a static one in a thermal bath at the Unruh temperature. As a result, the quantum system composed of two accelerated Unruh-DeWitt detectors will be driven into an asymptotic thermal state without entanglement when the separation between the two detectors is nonvanishing. This is what one expects based on the Unruh effect~\cite{Fulling1973,W. G. Unruh,Davies1975,Crispino2008}.

Secondly, from the perspective of entanglement generation,  the factor $\lambda$ which affects entanglement generation is independent of temperature for static detectors in a thermal bath, while it is related to acceleration  for accelerated detectors.

This shows that acceleration contributes to both the entanglement degradation and creation, while the  temperature of a thermal bath contributes to the entanglement degradation only. Therefore,  if the  entanglement creation effect overweighs the entanglement degradation effect, the maximum entanglement that can be generated during evolution may not decrease monotonically with the Unruh temperature as it does in the  thermal bath case. This explains why the anti-Unruh phenomenon in terms of the entanglement appears.

\section{Summary}
In this paper, we have investigated, in the framework of open quantum systems, the entanglement dynamics of a quantum system composed of two uniformly accelerated Unruh-DeWitt detectors coupled with fluctuating  massive scalar fields in the Minkowski vacuum. We first define a set of parameters, which play the roles of time delay, dissipation, decoherence and entanglement generation  in the evolution process. With the help of these parameters,  we study the influences of the mass of the field and the acceleration on the entanglement dynamics through comparing the corresponding results in the massless case and the thermal bath case respectively. The relevant conclusions are as follows.

Firstly, on the entanglement evolution rate, the entanglement evolution process for a quantum system coupled with massive fields is always slower compared with that of the one coupled with massless fields, which is advantageous to entanglement protection, but disadvantageous to  entanglement generation. Also, this time-delay effect brought about by the field being massive can however be counteracted by a large enough acceleration. It is interesting to note, however,  that in the thermal case, this time delay is not affected by the temperature. In particular, when the mass of the field is greater than the energy level spacing of the detectors, static detectors in a thermal bath will be locked up in its initial state, while entanglement generation is still possible in the acceleration case.

Secondly, on the entanglement degradation, we find that there are two effects causing entanglement degradation, namely the dissipation effect and decoherence effect. The accelerated detectors not only suffer  the same dissipation effect as that caused by a thermal bath, but also an additional decoherence effect, both of  which contribute to entanglement degradation.

Thirdly, on the entanglement creation, the region of spatial separation between the detectors within which entanglement can be generated is always enlarged for a static quantum system coupled with massive fields in a thermal bath compared with that in the massless case, while it can be both compressed and enlarged in the acceleration case.

Fourthly, on the asymptotic entanglement, two accelerated Unruh-DeWitt detectors  will be driven into an asymptotic thermal state without entanglement when the separation between the two detectors is nonvanishing, or  an asymptotic entangled state related to its initial state when the separation is vanishing, regardless of whether the mass of the field is greater or less than the energy level spacing of the detectors. However, for the thermal bath case, the detectors will be locked up in the initial state when the mass of the field is larger than the energy level spacing of the detectors, thus preserving the initial entanglement if it exists.

Finally, in conclusion, the entanglement dynamics for uniformly accelerated detectors coupled with  massive fields is essentially different from that of the static ones in a thermal bath at the Unruh temperature. In particular, the maximal concurrence of the quantum system generated during evolution may increase with acceleration when the acceleration is relatively small compared with the energy level spacing of the detectors  for any inter-detector separation, in contrast to the  monotonic decreases of the maximal concurrence with temperature for static detectors in a thermal bath, which can be considered as an anti-Unruh effect in terms of the entanglement generated.

\begin{acknowledgments}
This work was supported in part by the NSFC under Grants No. 11805063, No. 11690034, and No. 12075084, and the Hunan Provincial Natural Science Foundation of China under Grant No. 2020JJ3026.
\end{acknowledgments}

\appendix

\section{The expressions of the entanglement dynamic parameters in various cases}\label{various cases}

For the sake of simplicity, we assume that the coefficients $\chi^k_\mu$ defined in Eq. \eqref{Phi} satisfy $\sum^N_{k=1}\,\chi^k_\mu(\chi^k_\nu)^*=\delta_{\mu\nu}$. Then, $G_{ij}^{(\alpha\varrho)} (\Delta\tau)=\delta_{ij} G^{(\alpha\varrho)} (\Delta\tau)$, where $G^{(\alpha\varrho)} (\Delta\tau)$ is the standard Wightman function which can be expressed as
\bea\label{singleG}
G^{(\alpha\varrho)} (\Delta\tau)=\big\langle\phi\big(t_{\alpha}(\tau),\mathbf{x}_{\alpha}(\tau)\big)\phi\big(t_{\varrho}(\tau'),\mathbf{x}_{\varrho}(\tau')\big)\big\rangle.
\eea
In the free Minkowski spacetime, the scalar field operator $\phi(t,\mathbf{x})$ can be expanded as
\begin{eqnarray}\label{field operator}
\phi(t,\mathbf{x})=\int d^3 k\frac{1}{\sqrt{2\omega_k(2\pi^3)}}\big[a_k e^{i \mathbf{k} \mathbf{x}-i\omega_k t}+a^{\dagger}_k e^{-i \mathbf{k} \mathbf{x}+i\omega_k t}\big],
\end{eqnarray}
where $a_k$ and $a^{\dagger}_k$ are the annihilation and creation operators for  field quanta with  frequency $\omega_k$ and  momentum $k=|\mathbf{k}|$ satisfying the dispersion relation $\omega^2_k=k^2+m^2$, with $m$ being mass of the scalar fields. Substituting Eq.  \eqref{field operator} into Eq. \eqref{singleG}, we obtain
\begin{eqnarray}\label{C+chi2}
&&G^{(\alpha\varrho)}(\Delta\tau)={1\/4\pi^2}\int_{m}^{\infty}\frac{\sin{\left(\sqrt{\omega_k^2-m^2}|\Delta\vec{x}_{\alpha\varrho}|\right)}}{|\Delta\vec{x}_{\alpha\varrho}|}\left[2\langle N_{\omega_k}\rangle\cos{\left(\omega_k\Delta t_{\alpha\varrho}\right)}+e^{-i\omega_k\Delta t_{\alpha\varrho}}\right] d\omega_k,\hspace{0.8cm}
\end{eqnarray}
where  $|\Delta\vec{x}_{\alpha\varrho}|=\sqrt{(x_{\alpha}-x'_{\varrho})^2+(y_{\alpha}-y'_{\varrho})^2+(z_{\alpha}-z'_{\varrho})^2}$, $\Delta t_{\alpha\varrho}=t_{\alpha}-t'_{\varrho}$, and $N_{\omega_k}=a^{\dagger}_k a_k$ is the particle number operator.

\subsection{The acceleration case}
For a  uniformly accelerated quantum system composed of two Unruh-DeWitt detectors with a separation $L$ perpendicular to the acceleration, the trajectories of each detector are
\begin{eqnarray}\label{trajectories0}
&&t_{1}(\tau)=\frac{1}{a}\sinh{a\tau},\ \  x_{1}(\tau)=\frac{1}{a}\cosh{a\tau},\ \  y_{1}(\tau)=0,\ \  z_{1}(\tau)=0,\nonumber\\
&&t_{2}(\tau)=\frac{1}{a}\sinh{a\tau},\ \  x_{2}(\tau)=\frac{1}{a}\cosh{a\tau},\ \  y_{2}(\tau)=0,\ \  z_{2}(\tau)=L,
\end{eqnarray}
where $a$ is the proper acceleration of the detectors.
Substituting the trajectories above into Eqs. \eqref{C+chi2} and \eqref{D2}, and using the following integral formula
\begin{eqnarray}\label{BesselK}
%&&\int_{-\infty}^{+\infty}\cos{\left[(2\omega_k/a)\sinh{(a u/2)}\right]}e^{i\omega u} d u=(4/a)\cosh{(\pi\omega/a)K_{i2\omega/a}(2\omega_k/a)}\;,\nonumber\\
&&\int_{-\infty}^{+\infty}e^{-i(2\omega_k/a)\sinh{(a u/2)}}e^{i\omega u} d u=(4/a)e^{\pi\omega/a}{K_{i2\omega/a}(2\omega_k/a)}\;,
\end{eqnarray}
one  obtains
\begin{eqnarray}\label{C+chi acceleration}
&&{D_{\pm}^{(\alpha\varrho)}}={\varepsilon^2\omega\/\pi}\frac{2e^{\pm\pi\omega/a}}{\pi\omega/a}\int_{m\/a}^{\infty}{{\sin\big(a |\Delta z_{\alpha\varrho}|\sqrt{x^2-{m^2/{a^2}}}\big)}\/{a |\Delta z_{\alpha\varrho}|}}\;K_{i2\omega/a}\left(2x\right) dx,\nonumber\\
&&{D_{0}^{(\alpha\varrho)}}={\varepsilon^2 a\/\pi^2}\int_{m\/a}^{\infty}{{\sin\big(a |\Delta z_{\alpha\varrho}|\sqrt{x^2-{m^2/{a^2}}}\big)}\/{a |\Delta z_{\alpha\varrho}|}}\;K_{0}\left(2x\right) dx.
\end{eqnarray}
Here, $|\Delta z_{\alpha\varrho}|=L$ when $\alpha\neq\varrho$, and $|\Delta z_{\alpha\varrho}|=0$ when $\alpha=\varrho$. According to the definitions of  $\Omega,\;\eta$ and $\lambda$, which have been shown in Eqs. \eqref{Omegad}, \eqref{etad} and \eqref{lambdad} respectively, we can obtain that
\begin{eqnarray}
\Omega_a=\frac{\sinh\left(\pi\omega/a\right)}{\pi\omega/a}
\frac{m^2}{a^2}\left[K_{1+i\omega/a}\left(m/a\right)K_{-1+i\omega/a}\left(m/a\right)
-K_{i\omega/a}^2\left(m/a\right)\right],
\end{eqnarray}
\begin{eqnarray}
\eta_a=\coth\frac{\pi\omega}{a}
\end{eqnarray}
\begin{eqnarray}
\lambda_a=\frac{4a^2}{m^2}\frac{\int_{m\/a}^{\infty}{1\/{a L}}{\sin\big(a L\sqrt{x^2-{m^2/{a^2}}}\big)}\;K_{i2\omega/a}\left(2x\right) dx }{K_{1+i\omega/a}\left(m/a\right)K_{-1+i\omega/a}\left(m/a\right)
-K_{i\omega/a}^2\left(m/a\right)}.
\end{eqnarray}
Here, the subscript ``$_a$'' in $\Omega_a$, $\eta_a$ and $\lambda_a$ denotes the acceleration case.

\subsection{The thermal bath case}
For two static Unruh-DeWitt detectors in a thermal bath at temperature $T={1\/\beta}$, $\langle N_{\omega_k}\rangle=1/{(e^{\omega_{k}/{T}}-1)}=1/{(e^{\beta\omega_{k}}-1)}$.
% $|\Delta\vec{x}_{\alpha\varrho}|=|\Delta z_{\alpha\varrho}|$ and $\Delta t_{\alpha\varrho}=\Delta\tau$.
Plugging Eq. \eqref{C+chi2} into Eq. \eqref{D2}, we obtain
\begin{eqnarray}\label{C+chi thermal bath}
&&{D_{\pm}^{(\alpha\varrho)}}=
\begin{cases}
{\varepsilon^2\omega\/2\pi}\frac{\sin{\left(\sqrt{\omega^2-m^2}|\Delta z_{\alpha\varrho}|\right)}}{\omega |\Delta z_{\alpha\varrho}|}(\coth{\omega\/2T}\pm1),& \omega>m,\\
0,& 0\leq\omega\leq m,
\end{cases}
\end{eqnarray}
and
\begin{eqnarray}
&&{D_{0}^{(\alpha\varrho)}}=
\begin{cases}
\frac{\varepsilon ^2 T}{2\pi },& m=0,\\
0,& m\neq0.
\end{cases}
\end{eqnarray}
Then, direct calculations show that
\begin{eqnarray}
\Omega_\beta=
\begin{cases}
\sqrt{1-\frac{m^2}{\omega^2}},& \omega>m,\\
0,& \omega\leq m.
\end{cases},\;\;\;\;\;\eta_\beta=\coth\frac{\beta\omega}{2}=\coth\frac{\omega}{2T},\;\;\;\;\;\lambda_\beta=\frac{\sin(L\omega\Omega_\beta)}{L\omega\Omega_\beta}.
\end{eqnarray}
Here, the subscript ``$_{\beta}$'' in $\Omega_{\beta}$, $\eta_{\beta}$ and $\lambda_{\beta}$ denotes the thermal bath case.

\section{The limit properties of  $\Omega_a(m/\omega,a/\omega)$}\label{the properties Omega}
The factor $\Omega_a$ in Eq. \eqref{g}  can be obtained as follows,
\begin{eqnarray}\label{GG}
\Omega_a&=&\frac{4\sinh\left(\pi\omega/a\right)}{\pi\omega/a}\int_{0}^{\infty} \frac{k^2}{\sqrt{k^2+m^2/a^2}} K_{i2\omega/a}\left(2\sqrt{k^2+m^2/a^2}\right) dk\;\nonumber\\
&=&\frac{2\sinh\left(\pi\omega/a\right)}{\pi\omega/a}\int_{0}^{\infty} k K_{i\omega/a}^2\left(\sqrt{k^2+m^2/a^2}\right)dk\;\nonumber\\
&=&\frac{\sinh\left(\pi\omega/a\right)}{\pi\omega/a}
\frac{m^2}{a^2}\left[K_{1+i\omega/a}\left(m/a\right)K_{-1+i\omega/a}\left(m/a\right)
-K_{i\omega/a}^2\left(m/a\right)\right],
\end{eqnarray}
which can  be written as a function of dimensionless variables  as $\Omega_a(m/\omega,a/\omega)$.
Furthermore, it can be obtained that
\bea\label{app-c2}
\frac{d\Omega_a}{d m}=-\frac{2\sinh\left(\pi\omega/a\right)}{\pi\omega/a}\frac{m}{a^2} K_{i\omega/a}^2\left(m/a\right)\leq0.
\eea
In the following, we discuss the  limit properties of  $\Omega_a(m/\omega,a/\omega)$.

\subsection{The low-mass limit}
When $m\ll\omega$ and $m\ll a$, $\Omega_a(m/\omega,a/\omega)$ can be approximated by,
 \bea
\Omega_a(m/\omega,a/\omega)\approx1-{m^2\/2\omega^2}\bigg[1+\bigg(1+
{\omega^2\/a^2}\bigg)^{-{1\/2}}\cos\bigg(2{\omega\/a}\ln{m\/2a}-\varphi({\omega/a})\bigg)\bigg],
\label{low mass and high acceleration limit}
\eea
 where $\varphi(\alpha)$ is an argument function defined as
\bea
\varphi(\alpha)=\arg[{(1+i\alpha)\Gamma^2(i\alpha)}]\;.
\eea
Note that  $\Gamma(z)$ is the Euler gamma function. This shows that the mass-dependent term gives a small correction
which is negative. When the mass of the scalar field approaches 0,
\bea\lim_{m\rightarrow0}\Omega_a(m/\omega,a/\omega)=1,\eea and the result reduces to that in the massless case as expected.
\subsection{The high-mass limit}
When the mass of scalar field is much larger than the energy level
spacing and the acceleration, i.e. $m\gg\omega$ and $m\gg a$, $\Omega_a(m/\omega,a/\omega)$ can be approximated by,
\bea\label{Omegalimitmh1}
\Omega_a(m/\omega,a/\omega)\approx
e^{-2{m\/a}}{\sinh(\pi\omega/a)\/2\omega/a}\;.
\eea
For a fixed acceleration, it exponentially approaches to zero as the mass of the field increases. Thus, \bea\lim_{m\rightarrow\infty}\Omega_a(m/\omega,a/\omega)=0.\eea

\subsection{The low-acceleration limit}
In the low-acceleration limit, i.e. the
acceleration  is much smaller than the energy level
spacing and the field mass, $a\ll\omega$ and $a\ll m$, $\Omega_a(m/\omega,a/\omega)$ can be approximately written as (for a detailed
derivation see Ref. \cite{Y. B. Zhou})
\bea\label{Omegalimita}
\Omega_a(m/\omega,a/\omega)\approx\left\{
                                     \begin{array}{ll}
                                       \sqrt{1-({m/\omega})^2}\big[1+\cos(2A_1\omega/a)\big({1\/72}+B\big){a\/A_1\,\omega}\big],\;\; \;\; & \omega> m\;,\\
                                       {1\/2}\sqrt{({m/\omega})^2-1}\,\big({25\/72}+B\big)\,{a\/A_2\,\omega}\;\;e^{-{2\,A_2\,\omega/a}}, & \omega\leq m\;.
                                     \end{array}
                                   \right.\label{f}
                                    \eea
Here, $A_1$, $A_2$ and $B$ are positive, which are defined
respectively as
\bea\label{ABC}
&&A_1=\ln\frac{\omega+\displaystyle\sqrt{\omega^2-m^2}}{m}-\displaystyle\sqrt{1-(m/\omega)^2}\;,
\nn\\&&A_2=\displaystyle\sqrt{({m/\omega})^2-1}-\arccos({\omega}/{m})\;,\nonumber\\&&
B=\left(1+C^2/3\right)A C/4\;,
 \eea
where
$C=1/\sqrt{1-m^2/\omega^2}$ and $A=A_1$, for
$\omega> m$;  and
$C=1/\sqrt{m^2/\omega^2-1}$ and $A=A_2$, for
$\omega\leq m$. So, when the acceleration $a$ approaches zero,
\bea
\lim_{a\rightarrow0}\Omega_a(m/\omega,a/\omega)=\left\{
                                     \begin{array}{ll}
                                       \sqrt{1-({m/\omega})^2}, \quad\;\quad\;\;\;& \omega> m,\\
                                       0, & \omega\leq m,
                                     \end{array}
                                   \right.\eea
which is the results of a static quantum system in vacuum \cite{Zhou2020}, as expected.
\subsection{The high-acceleration limit}
For the high-acceleration limit where
$a\gg\omega$ and $a\gg m$, keeping only the lowest-order correction term,
one finds that
 \bea
\Omega_a(m/\omega,a/\omega)\approx1-{m^2\/a^2}\bigg[\bigg(\ln{m\/a}\bigg)^2-1.23\ln{m\/a}+0.63\bigg]\;,\label{High
acceleration limit}
 \eea
which gives
\bea\label{Omegalimita2}
\lim_{a\rightarrow\infty}\Omega_a(m/\omega,a/\omega)=1.
\eea

\section{The limit properties of $ \lambda_a(m/\omega,a/\omega,L\omega) $}\label{the properties lambda}
For the acceleration case, the factor $\lambda$ defined in Eq. \eqref{lambdad} can be expressed in the  following  form
\begin{eqnarray}\label{lambdag}
\lambda_{a}&=&\frac{\int_{0}^{\infty} \frac{k}{\sqrt{k^2+m^2/a^2}} K_{i2\omega/a}\left(2\sqrt{k^2+m^2/a^2}\right)\frac{\sin\left(a L k\right)}{a L} dk}{\int_{0}^{\infty} \frac{k^2}{\sqrt{k^2+m^2/a^2}} K_{i2\omega/a}\left(2\sqrt{k^2+m^2/a^2}\right)dk}\;\nonumber\\
&=&\frac{\int_{0}^{\infty} k K_{i\omega/a}^2\left(\sqrt{k^2+m^2/a^2}\right)J_0\left(a L k\right) dk}{\int_{0}^{\infty} k K_{i\omega/a}^2\left(\sqrt{k^2+m^2/a^2}\right)dk}\;\nonumber\\
&=&\frac{4a^2}{m^2}\frac{\int_{m\/a}^{\infty}{1\/{a L}}{\sin\big(a L\sqrt{x^2-{m^2/{a^2}}}\big)}\;K_{i2\omega/a}\left(2x\right) dx }{K_{1+i\omega/a}\left(m/a\right)K_{-1+i\omega/a}\left(m/a\right)
-K_{i\omega/a}^2\left(m/a\right)},
\end{eqnarray}
where $J_{\nu}(x)$ and $K_{\nu}(x)$ are the Bessel function of the first type  and the  modified Bessel function of the second type, respectively.
The factor $\lambda_a$ can  be written as a function of dimensionless variables as $\lambda_a(m/\omega,a/\omega,L\omega)$.

\subsection{The low-mass limit}
Keeping only the lowest-order term, in the low-mass limit where
$m\ll\omega$ and $m\ll a$, the limit of $\lambda_a$ when $m\rightarrow0$ is
\bea
\lim\limits_{m\rightarrow0}\lambda_a(m/\omega,a/\omega,L\omega)=\frac{\sin\left(\frac{2\omega}{a}\sinh^{-1}\frac{a L}{2}\right)}{L\omega\sqrt{1+a^2 L^2/4}}\;,
\eea
which reduces to the result in the massless case.

\subsection{The high-mass limit}
When $m\gg a$ and $m\gg\omega$,  $\lambda_a$ can be approximated as
\begin{eqnarray}\label{limitm1}
\lambda_a(m/\omega,a/\omega,L\omega)\approx\frac{e^{-(\sqrt{4+a^2L^2}-2)\frac{m}{a}}}{1+a^2L^2/4}.
\end{eqnarray}
Therefore,  in the limit of $m\rightarrow\infty$, $\lambda_a$  is dependent on the value of $a L$.
\begin{eqnarray}\label{limitm2}
\lim\limits_{m\rightarrow\infty}\lambda_a(m/\omega,a/\omega,L\omega)=
\begin{cases}
1,& a L=0,\\
0,& a L\neq0.
\end{cases}
\end{eqnarray}

\subsection{The low-acceleration limit}
If $a\ll m$, $a\ll\omega$ and $a\ll 1/L$,
\begin{eqnarray}\label{limita}
\lim\limits_{a\rightarrow0}\lambda_a(m/\omega,a/\omega,L\omega)=
\begin{cases}
\frac{\sin\left(L \sqrt{\omega^2-m^2}\right)}{L \sqrt{\omega^2-m^2}},& m/\omega<1,\\
1,& m/\omega\geq1,
\end{cases}\;
\end{eqnarray}
which reduces to the result in the vacuum case Eq. \eqref{lambda b}.
\subsection{The high-acceleration limit}
When $a\gg m$,  $\lambda_a$ can be approximated as
\begin{eqnarray}\label{limitaggm}
\lambda_a(m/\omega,a/\omega,L\omega)\approx\frac{\sin\left(\frac{2\omega}{a}\sinh^{-1}\frac{a L}{2}\right)}{L\omega\sqrt{1+a^2 L^2/4}},
\end{eqnarray}
which is the result in the massless case. Based on this, if $a\gg\omega$, it can further be approximate as,
\begin{eqnarray}\label{limitaggm aggo}
\lambda_a(m/\omega,a/\omega,L\omega)\approx\frac{4 \sinh ^{-1}\left(\frac{a L}{2}\right)}{a L \sqrt{a^2L^2+4}}.
\end{eqnarray}
Furthermore, if $a\gg1/L$, then
\begin{eqnarray}\label{limitaggm aggo aggL}
\lambda_a(m/\omega,a/\omega,L\omega)\approx \frac{4 \log (a L)}{a^2L^2}.
\end{eqnarray}
Here, if $a L\ll 1$,  the approximate expression of $\lambda_a$ can be written as $\lambda_a\approx 1-a^2L^2/6$. Thus, we can get the limit of $\lambda_a$ when $a\to\infty$ as
\begin{eqnarray}\label{hlimita}
\lim\limits_{a\rightarrow\infty}\lambda_a(m/\omega,a/\omega,L\omega)=
\begin{cases}
1,& L=0,\\
0,& L\neq0.
\end{cases}
\end{eqnarray}

\section{The limit properties of $ \zeta^{\pm}_a(m/\omega,a/\omega,L\omega) $}\label{the properties zeta}
For the acceleration case, the factors $\zeta^{\pm}$ defined in Eq. \eqref{zeta+gamma d} can be expressed in the  following  form
\begin{eqnarray}\label{zeta+-d}
\zeta^{\pm}_a=\frac{{m^2}\left[K_{1}^2\left(m/a\right)-K_{0}^2\left(m/a\right)\right]\pm 4{a^2}\int_{m\/a}^{\infty}{\frac{\sin\big(a L\sqrt{x^2-{m^2/{a^2}}}\big)}{a L}}\;K_{0}\left(2x\right) dx }{2{m^2}\left[K_{1+i\omega/a}\left(m/a\right)K_{-1+i\omega/a}\left(m/a\right)
-K_{i\omega/a}^2\left(m/a\right)\right]\cosh\left(\pi\omega/a\right)},
\end{eqnarray}
where $K_{\nu}(x)$ is the modified Bessel function of the second type.
The factor $\zeta^{\pm}_a$ can be written as a function of dimensionless variables as $\zeta^{\pm}_a(m/\omega,a/\omega,L\omega)$.

\subsection{The low-mass limit}
Keeping only the lowest-order term, in the low-mass limit where
$m\ll\omega$ and $m\ll a$, the limit of $\zeta^{\pm}_a$ when $m\rightarrow0$ is
\bea
\lim\limits_{m\rightarrow0}\zeta^{\pm}_a(m/\omega,a/\omega,L\omega)=\frac{\tanh\left(\pi\omega/a\right)}{\pi\omega/a}\frac{1}{2}\left[1\pm\frac{2\sinh^{-1}\left({a L}/{2}\right)}{a L\sqrt{1+a^2L^2/4}}\right]\;,
\eea
which reduces to the result in the massless case.

\subsection{The high-mass limit}
When $m\gg a$ and $m\gg\omega$, with the help of Eqs. \eqref{Omegalimitmh1} and \eqref{limitm1}, $\zeta^{\pm}_a$ can be approximated as
\begin{eqnarray}\label{zetalimitm1}
\zeta^{\pm}_a(m/\omega,a/\omega,L\omega)\approx\frac{1}{2\cosh(\pi\omega/a)}\left[1\pm\frac{e^{-(\sqrt{4+a^2L^2}-2)\frac{m}{a}}}{1+a^2L^2/4}\right].
\end{eqnarray}
Therefore,  in the limit of $m\rightarrow\infty$, $\zeta^{\pm}_a$  is dependent on the value of $L$.
\begin{eqnarray}\label{zetalimitm2}
&&\lim\limits_{m\rightarrow\infty}\zeta^{+}_a(m/\omega,a/\omega,L\omega)=
\begin{cases}
{\text{sech}(\pi\omega/a)},& L=0,\\
\frac{1}{2}\text{sech}(\pi\omega/a),& L\neq0,
\end{cases}\\
&&\lim\limits_{m\rightarrow\infty}\zeta^{-}_a(m/\omega,a/\omega,L\omega)=
\begin{cases}
0,& L=0,\\
\frac{1}{2}\text{sech}(\pi\omega/a),& L\neq0.
\end{cases}
\end{eqnarray}

\subsection{The low-acceleration limit}
If $a\ll m$, $a\ll\omega$ and $a\ll 1/L$,
with the help of Eqs. \eqref{Omegalimita} and \eqref{limita}, one can find  that
\begin{eqnarray}\label{zetalimita1}
\zeta^{\pm}_a(m/\omega,a/\omega,L\omega)\approx
\begin{cases}
\frac{1}{2}\frac{\tanh\left(\pi\omega/a\right)}{\pi\omega/a}(1\pm1),& m/\omega=0,\\
\frac{\pi}{4}\frac{\tanh\left(\pi\omega/a\right)}{\pi\omega/a}\frac{e^{-2m/a}}{\sqrt{1-m^2/\omega^2}}(1\pm1),& 0<m/\omega<1,\\
\frac{\pi}{8}\frac{\tanh\left(\pi\omega/a\right)}{\pi\omega/a}\frac{A_2\omega}{a}\frac{e^{-2\left({m}-\omega A_2\right)/a}}{\sqrt{m^2/\omega^2-1}\left({25\/72}+B\right)}(1\pm1),& m/\omega\geq1,
\end{cases}\;
\end{eqnarray}
Here, $A_2$ and $B$ have been defined in Eq. \eqref{ABC}. Then, from the approximate expression \eqref{zetalimita1}, one can obtain that
\begin{eqnarray}\label{zetalimita2}
\lim\limits_{a\rightarrow0}\zeta^{\pm}_{a}(m/\omega,a/\omega,L\omega)=0.
\end{eqnarray}

\subsection{The high-acceleration limit}
When $a\gg m$, $a\gg\omega$, taking Eqs. \eqref{Omegalimita2} and \eqref{hlimita} into  Eq. \eqref{zeta+gamma re}, we can get the limit of $\zeta^{\pm}_a$ when $a\to\infty$ as
\begin{eqnarray}\label{zetalimita2}
&&\lim\limits_{a\rightarrow\infty}\zeta^{+}_a(m/\omega,a/\omega,L\omega)=
\begin{cases}
1,& L=0,\\
\frac{1}{2},& L\neq0.
\end{cases}\\
&&\lim\limits_{a\rightarrow\infty}\zeta^{-}_a(m/\omega,a/\omega,L\omega)=
\begin{cases}
0,& L=0,\\
\frac{1}{2},& L\neq0.
\end{cases}
\end{eqnarray}

\end{document}